\documentclass[twocolumn, superscriptaddress,preprintnumbers,amsmath,amssymb,longbibliography]{revtex4-1}
\usepackage{graphicx}
\usepackage{mathtools}
\usepackage{microtype}
\usepackage{amsmath}

\usepackage{epstopdf}
\usepackage{color,soul}

\begin{document}

\preprint{Version: \today}

\title{Intermodulation in Nonlinear SQUID Metamaterials: Experiment and Theory}

\author{Daimeng Zhang}
\affiliation{Department of Electrical and Computer Engineering, University of Maryland, College Park, Maryland 20742-3285, USA}
\affiliation{Center for Nanophysics and Advanced Materials, University of Maryland, College Park, Maryland 20742-4111, USA}
\author{Melissa Trepanier}
\affiliation{Center for Nanophysics and Advanced Materials, University of Maryland, College Park, Maryland 20742-4111, USA}
\affiliation{Department of Physics, University of Maryland, College Park, Maryland 20742-4111, USA}
\author{Thomas Antonsen}
\affiliation{Department of Electrical and Computer Engineering, University of Maryland, College Park, Maryland 20742-3285, USA}
\affiliation{Department of Physics, University of Maryland, College Park, Maryland 20742-4111, USA}
\author{Edward Ott}
\affiliation{Department of Electrical and Computer Engineering, University of Maryland, College Park, Maryland 20742-3285, USA}
\affiliation{Department of Physics, University of Maryland, College Park, Maryland 20742-4111, USA}
\author{Steven M. Anlage}
\affiliation{Department of Electrical and Computer Engineering, University of Maryland, College Park, Maryland 20742-3285, USA}
\affiliation{Center for Nanophysics and Advanced Materials, University of Maryland, College Park, Maryland 20742-4111, USA}
\affiliation{Department of Physics, University of Maryland, College Park, Maryland 20742-4111, USA}

\date{\today}

\begin{abstract}
The response of nonlinear metamaterials and superconducting electronics to two-tone excitation is critical for understanding their use as low-noise amplifiers and tunable filters. A new setting for such studies is that of metamaterials made of radio frequency superconducting quantum interference devices (rf-SQUIDs). The two-tone response of self-resonant rf-SQUID meta-atoms and metamaterials is studied here via intermodulation (IM) measurement over a broad range of tone frequencies and tone powers. A sharp onset followed by a surprising strongly suppressed IM region near the resonance is observed. 
Using a two time scale analysis technique, we present an analytical theory that successfully explains our experimental observations. 
The theory predicts that the IM can be manipulated with tone power, center frequency, frequency difference between the two tones, and temperature. This quantitative understanding potentially allows for the design of rf-SQUID metamaterials with either very low or very high IM response.

\end{abstract}

\maketitle

\section{Introduction}
Nonlinearity is a key consideration in a wide range of important applications including amplifiers \cite{Vijay2009, Eom2012, Siddiqi2013} and tunable filters \cite{Mateu2007}. Introduction of nonlinearity into metamaterials facilitates tunability, design flexibility, and self-induced nonlinear responses \cite{Lapine2014, Shadrivov2008}, giving rise to developments in metamaterial-based amplifiers \cite{Castellanos2008, Lee2013} , filters \cite{Gil2005, Meta_filters_review2008, Watts2016} and antennas \cite{Lim2005, Ziolkowski2006, Dong2012}. However, as data streams containing multi-frequency signals pass through these nonlinear components, they generate intermodulation (IM) products via frequency mixing \cite{pedro2002intermodulation}. The same issue appears in intrinsically nonlinear superconducting electronics. The IM between two input frequencies $f_1$ and $f_2$ leads to products at frequencies $pf_1\pm qf_2$ ($p$ and $q$ are integers), forming side bands and additional noise that could diminish the performance of superconducting devices \cite{shen1994high,  Abuelma'atti1993, Willemsen1997, Dahm1997JAP, McDonald1998, Hammond1998, Benz1999, Remillard2003, Mateu2003, Mateu2007, Mateu2009, Rocas2011, Tholén2014}. On the other hand, IM generation can be used as a diagnostic to determine various types of defects in superconductors \cite{Oates2001, Oates2003, Zhuravel2003,Xin2002}, to study unconventional superconductors \cite{Willemsen1998, Hao2001, Oates2001, Xin2002,Oates2003, Lamura2003, Velichko2004, Oates2005, Oates2007, Jang2009, Agassi2009, Pease2010, Agassi2014}, and to amplify microwave signals \cite{Abdo2009, Tholén2014, Abdo2006, Eom2012}, even at the quantum limit in Josephson parametric amplifiers \cite{Vijay2009, Siddiqi2013} and Josephson metamaterials \cite{Castellanos2008}. Therefore, IM is of mutual research interest in wireless communication, nonlinear metamaterials, as well as in quantum information processing, and superconducting electronics and materials. Extensive measurement and theory have been devoted to IM in these fields \cite{Dahm1996, Sollner1996, Willemsen1997, Dahm1997theoryres, Willemsen1999, Dahm1999, Dahm1999JAP, Hu1999, Vopilkin2000, Hutter2010_reconstructIMD, Mateu2009}.

Rf-SQUID metamaterials combine the advantages of superconducting electronics and nonlinear metamaterials \cite{Lapine2014, Anlage2011,Jung2014review}.  An rf-SQUID is the macroscopic quantum version of a split ring resonator (SRR) with the gap capacitance in the SRR replaced by a nonlinear Josephson junction. SQUIDs can be very sensitive to dc and rf magnetic flux, on the scale of the flux quantum $\Phi_0=h/2e=2.07\times10^{-15}$ Tm$^2$, where $h$ is Planck's constant and $e$ is the elementary charge.
 Previous work reveals that rf-SQUID meta-atoms and metamaterials have a resonant frequency tunability of up to $80$THz/Gauss by varying the dc magnetic flux when the driving rf flux amplitude is low \cite{Jung2013, Butz20132, Trepanier2013}. In Ref. \cite{Daimeng2015} the authors studied the bistability of rf-SQUID meta-atoms and metamaterials driven by intermediate rf flux amplitudes. The bistability results in a lower resonant frequency and a nearly full disappearance of resonance absorption (transparency). Such broadband transparency can be switched on and off via drive frequency, signal amplitude, or dc flux hysteresis \cite{Daimeng2015}.
These properties make rf-SQUID metamaterials attractive for tunable filters, gain-modulated antennas \cite{Mukhanov2014}, and wideband power limiters for direct-digitizing rf receivers \cite{Mukhanov2008} in next-generation wireless communication systems. 

Basically, an rf-SQUID is a nonlinear resonator with a manipulatable resonant frequency and absorption that depend on the dc and rf flux amplitudes, the temperature, and the drive signal history \cite{Du2006, Lazarides2007, Maimistov2010,  Lazarides2013, Jung2013, Butz20132, Trepanier2013, Jung2014, Tsironis2014,Jungthesis,Daimeng2015,Tsironis2016}. We will study IM generation around this tunable, bi-stable resonance.

In this paper we report comprehensive results from experimental and theoretical IM studies of rf-SQUID meta-atoms and metamaterials around resonance. We focus on the case where two input signals have the same amplitude, as opposed to IM amplification experiments where one tone is much stronger than the other. We find that under certain combinations of tone power and frequency, the SQUID shows a sudden onset of the $3^{rd}$ order IM generation followed by a near-zero $3^{rd}$ order IM generation (gap). This phenomenon is a result of the bi-stable properties of rf SQUIDs. This intrinsic suppression of IM generation may be useful as a mechanism for depressing signal mixing in  communication applications. A detailed theoretical model is presented to explain this surprising gap feature in IM generation. The intensity of IM generation sensitively depends on the parameters of the rf-SQUIDs, and can be modulated by dc/rf magnetic field, and temperature, potentially allowing one to design and tune the IM generation to meet various requirements for applications. 


\section{Experiment Details}
\begin{figure}[]
\centering
\includegraphics[width=85mm]{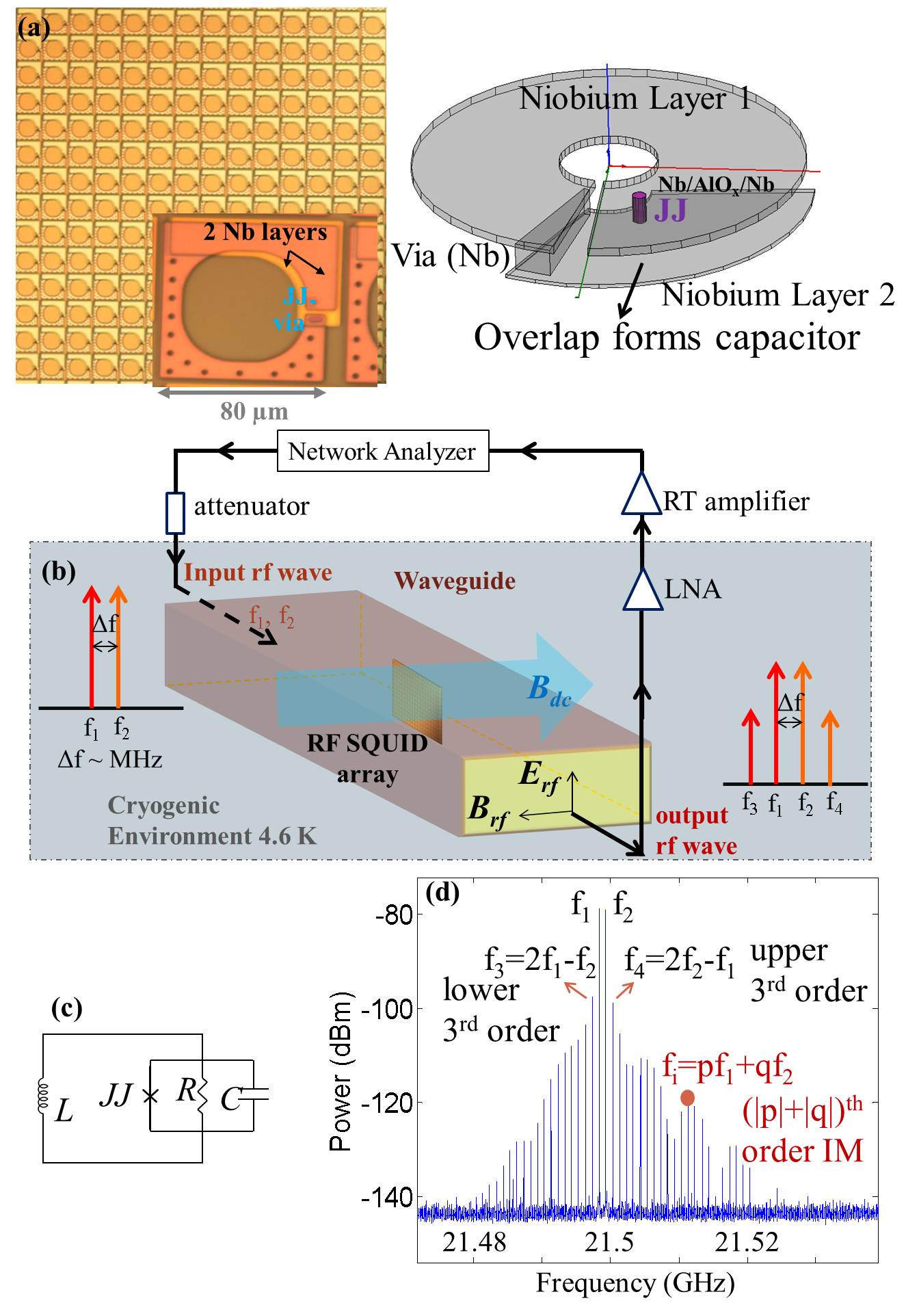}
\caption{(a) Left: The optical image of meta-atoms of a $27\times27$ array metamaterial. Inset shows details of a single SQUID. Right: The 3D structure of a single rf-SQUID. The distance between two Niobium layers is exaggerated to show the overlap capacitance. (b) The experimental setup for our IM measurements. (c) The circuit model for a single SQUID. (d) Experimental measurements of output power from the 27$\times$27 rf SQUID metamaterial at a temperature of $T= 4.6$ K as a function of frequency when two signals of the same amplitude are injected at a center frequency of 21.499 GHz and a difference frequency of 1 MHz.}
\label{IMD_exp}
\end{figure}

Two dimensional metamaterials were constructed by positioning rf SQUID meta-atoms in a square grid array on a planar substrate (Fig. \ref{IMD_exp} (a)). The single rf SQUID meta-atoms, and the metamaterials, were fabricated using the Hypres 0.3 $\mu $A$/ \mu $m$^2$ Nb/AlO$_x$/Nb junction process on silicon substrates, and the meta-atom has a superconducting transition temperature $T_c = 9.2$ K. A 3D perspective drawing of a single rf-SQUID is shown in Fig. \ref{IMD_exp} (a) . Two Nb films (135 nm and 300 nm thick) connected by a via and a Josephson junction make up the superconducting loop with geometrical inductance $L$. The capacitance $C$ has two parts: the overlap between two layers of Nb with 200 nm thick SiO$_2$ dielectric in between, and the Josephson junction intrinsic capacitance. The rf SQUIDs are designed to be low-noise
($\Gamma = {2 \pi k_B T}/{(\Phi_0 I_c)}<1$ where $T$ is the temperature , $I_c$ is the critical current in the Josephson junction, $\Phi_0=h/2e$ is the quantum flux, and
$L_F = ({k_B T})^{-1}[{\Phi_0}/({2 \pi})]^2>> L$ \cite{Chesca1998})
and non-hysteretic ($\beta_{rf}={2 \pi L I_c}/{\Phi_0}<1$). No dc magnetic flux is applied for this set of experiments. 

In the experimental setup Fig. \ref{IMD_exp} (b), the rf-SQUID array sits in a rectangular waveguide orientated so that the rf magnetic field of the TE mode is perpendicular to the rf-SQUIDs. Before each two-tone experiment, a single-tone transmission experiment is conducted to determine the resonant frequency at which the system has maximum power absorption. 
IM products are then measured systematically around the resonance; two signals of frequencies $f_1$ and $f_2$ having the same amplitude and a small difference in frequency $\Delta f=f_2-f_1>0$ are injected. The output signal contains the two main tones and their harmonics, as well as IM products.

An example of the generation of an IM spectrum in the metamaterial around resonance (of a 27$\times$27 array of rf SQUIDs) is shown in Fig. \ref{IMD_exp} (d) with $\Delta f=1$MHz. This spectrum was measured under a fixed tone center frequency and a fixed tone power. The output signal at frequency $f_i=pf_1+qf_2$ is called the $(|p|+|q|)^{th}$ order IM. We focus on nearby IM products which are of the $3^{rd}, 5^{th}, 7^{th}, ...$ order. The IM signals generated at nearby frequencies $f_3=2f_1-f_2$ and $f_4=2f_2-f_1$, called the lower and upper $3^{rd}$ order IM ($f_2>f_1$), respectively, are of most concern in communications and mixing applications. When the metamaterial is superconducting (measured at $T=4.6$ K), there is strong IM generation observed above the noise floor up to $51^{st}$ order. There is no observed IM output when temperature is above the transition temperature, $T_c=$ 9.2 K.  

\begin{figure}[]
\centering
\includegraphics[width=85mm]{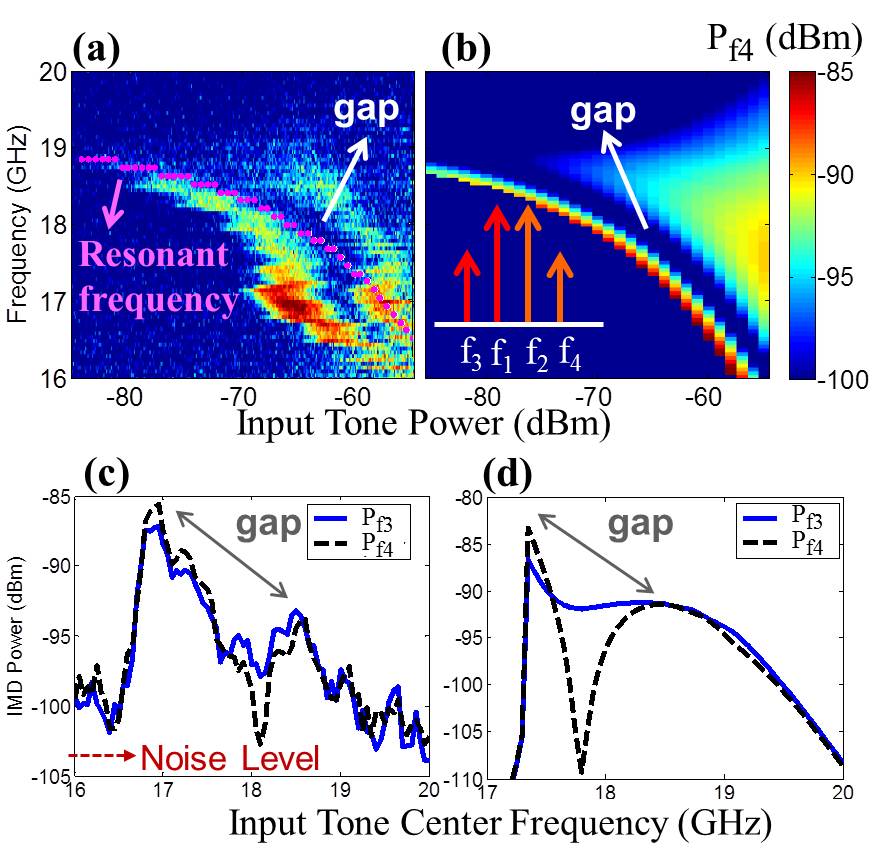}
\caption{The upper $3^{rd}$ IM power $P_{f_4}$ generated from a single rf-SQUID meta-atom as a function of the applied rf flux and the center frequency of the two tones for (a) experiment and (b) numerical simulation. The purple curve indicates the resonant frequency for a single-tone excitation. The frequency cut for output power at the third IM $P_{f_3}$ (blue solid line) and $P_{f_4}$ (black dashed line) at -65 dBm for (c) experiment and (d) simulation. The spacing between the two input tones is 10 MHz, and the temperature is 4.6 K. }
\label{IMD_asym}
\end{figure}

The IM spectrum changes considerably as the center frequency and tone power are varied. We mainly examine the modulation of the $3^{rd}$ order IM power. Again we first search for resonance in a single-tone experiment as the input power varies. In the intermediate power regime, higher input power results in a shift of the resonant frequency to lower values \cite{Daimeng2015}, as seen in the purple curve in Fig. \ref{IMD_asym} (a). The $3^{rd}$ order IM power is then measured with two-tone input around the resonance.
Fig. \ref{IMD_asym} (a) shows the upper $3^{rd}$ order IM power $P_{f_4}$ (colors) generated from a single rf-SQUID meta-atom as a function of the input tone power (horizontal axis) and the center frequency (vertical axis) of the two tones. The IM generation generally follows the resonant frequency curve. Intermodulation is small for low input tone powers ($<-80$ dBm), with a peak just below the resonant frequency. As the input power increases, the IM generation also increases while shifting to lower frequencies. At the same time a second peak appears above the resonant frequency, forming an IM gap where the IM is reduced to nearly the noise level around the resonant frequency. The same phenomenon is observed for a 7$\times$7 array rf-SQUID metamaterial and an 11$\times$11 array rf-SQUID metamaterial. Operating the meta-atom or metamaterial in the gap regime minimizes the $3^{rd}$ order IM frequency mixing. 

Figure \ref{IMD_asym} (c) compares the measured lower and upper $3^{rd}$ order IM products ($P_{f_3}$ and $P_{f_4}$) as a function of frequency around the gap feature at -65 dBm. Both IM powers show a sharp onset above the noise level at around 17 GHz, and decrease to a minimum value at 18 GHz, then reach another peak at around 18.4 GHz before dropping continuously at higher frequencies. However, the upper tone $P_{f_4}$ has a higher peak and a substantially lower dip than the lower tone $P_{f_3}$. This asymmetry between two same-order IM tones was also observed in other SQUID samples and in our numerical simulations. We now wish to explore the origins of the features seen in the data, including the sharp onset and the dip in the $3^{rd}$ IM generation, as well as the asymmetry between the upper and lower IM output signals.

\section{Modeling}
\subsection{Numerical Simulation}
In this section we explore a simple circuit model that reproduces the effects seen in the previous sections. A single rf SQUID can be treated as a Resistively and Capacitively Shunted Josephson Junction (RCSJ-model) in parallel with superconducting loop inductance (Fig. \ref{IMD_exp} (c)). We assume a uniformly driven and uncoupled SQUID array metamaterial can also be described by the single junction RCSJ-model. The macroscopic quantum gauge-invariant phase difference across the junction $\delta$ determines the current through the junction $I=I_c\sin{\delta}$ ($I_c$ is the critical current of the junction). In a closed superconducting loop $\delta$ is related to the total magnetic flux inside the loop: $\delta - 2\pi \Phi_{tot} / \Phi_0 = 2 \pi n$, where $n$ is an integer, and again $\Phi_0=h/2e$. Here we can take $n$ to be $0$ without loss of generality as shifting $\delta$ by $2\pi$ leaves the current $I$ unchanged \cite{Likharev1986}. 
The voltage across the junction 
can be written as $V= 2\pi\Phi_0 d\delta / dt$.  

The time evolution of the phases is determined by the RCSJ circuit equation \cite{Likharev1986}, obtained by demanding that the total flux through the loop $\Phi_{tot}$ is the combination of the dc and rf applied flux ($\Phi_{dc}+\Phi_{rf}(t)$), and the induced flux due to the self inductance $L$ of the loop,
\begin{equation}
\label{flux}
\Phi_{tot}=\Phi_{dc}+\Phi_{rf}(t)-L(I_c \sin{\delta} + \frac{V}{R} + C\frac{dV}{dt}).
\end{equation}
Here, $I_c\sin{\delta}+V/R+CdV/dt $ is the total current through the loop, which flows through the parallel combination of the junction, shunt resistance $R$ and capacitance $C$ in the RCSJ model. Replacing $\Phi_{tot}$ by $\Phi_0 \delta/2\pi $ and $V$ by $\Phi_0 d\delta / dt$ in Eq. (\ref{flux}) and rearranging terms, we obtain the dimensionless RCSJ equation:

\begin{equation}
\label{delta_IMD}
\frac{d^2\delta}{d\tau^2} + \frac{1}{Q} \frac{d\delta}{d\tau} +\delta+ \beta_{rf} \sin{\delta}\\
=\phi_{dc} + \phi_{rf}(\tau)
\end{equation}
where $\beta_{rf}=2\pi L I_c/\Phi_0$, $\phi_{dc}=2\pi \Phi_{dc}/\Phi_0$, $\phi_{rf}= 2\pi \Phi_{rf}/\Phi_0$, $\omega_{geo}=(LC)^{-1/2}$,   $\tau=\omega_{geo}t$, and $Q=R\sqrt{C/L}$.



Typical parameter values are as follows. The inductance, $L=280$ pH, of the single SQUID meta-atom is calculated numerically by Fasthenry based on its geometrical structure \cite{Fasthenry}. Other parameters such as the capacitance, $C=0.495$ pF, the shunt resistance in the junction, $R=1780$ Ohm (4.6 K), and the critical current, $I_c=1.15 \mu$A, are determined by fitting to the measured geometrical resonant frequency $\omega_{geo}/2\pi=13.52$ GHz, the measured quality factor $Q=75$, and the quantity $\beta_{rf}=0.98$. The quantities $\omega_{geo}$, $Q$, and $\beta_{rf}$ were directly measured in previous single-tone transmission experiments \cite{Trepanier2013, Daimeng2015}. For our setup, the rf flux $\phi_{rf}$ driving the loop results from the injected rf power inside the rectangular waveguide. Note that the single SQUID meta-atom has an inner diameter of $200 \mu$m, and an outer diameter of $800 \mu$m. Other meta-atoms in our SQUID metamaterials all have smaller sizes. Thus the rf flux amplitude through the SQUID loop is always much smaller than the flux quantum in the rf power range we consider in this work. Thus, $|\phi_{rf}|<2\pi$.

The time-dependent functional form of the rf flux is determined by the driving signal. To study intermodulation, the circuit is driven with two tones, which generally can be written
\begin{equation}
\phi_{rf}=\phi_{rf,1} \sin(\Omega_1 \tau + \theta_1)+\phi_{rf,2} \sin(\Omega_2 \tau + \theta_2)
\end{equation}
where $\Omega_{1,2}=2\pi f_{1,2}/\omega_{geo}$ and $f_1$ and $f_2$ are the frequencies of the two injected signals. Here the two tones have different amplitudes $ \phi_{rf,1} $ and $ \phi_{rf,1} $, and phases $\theta_1$ and $\theta_2$.

The driving flux can also be written in the form of a complex phasor envelope modulated by a carrier at the mean frequency $\Omega=(\Omega_1+\Omega_2)/2$,

\begin{equation}
\label{phasor}
\phi_{rf,a}=Re[e^{i\Omega \tau-i\pi/2}\phi_e(\tau)]
\end{equation}
where the envelope function $\phi_e(\tau)=\phi_{rf,1} \exp(-i\Delta\Omega \tau/2+i\theta_1)+\phi_{rf,2} \exp(i\Delta\Omega \tau/2+i\theta_2) $ and $\Delta\Omega=\Omega_2-\Omega_1>0$ is the difference frequency. For the situation in our experiment,  $\Delta\Omega<<\Omega$, $i.e.$, the carrier frequency is much greater than the envelope frequency. This will lead to a number of simplifications in the analysis. At present it allows us to argue that the results will not depend on the relationship between the carrier and the envelope phases. Since the relative phase between the carrier and the envelope is unimportant we may shift the time axis in the carrier and the envelope independently. Shifting time in the carrier by $\tau_{sc}=-\Omega^{-1}(\theta_1+\theta_2)$ and in the envelope by $\tau_{se}=\Delta\Omega^{-1}(\theta_1-\theta_2)$ removes the phases $ \theta_1 $
 and $ \theta_2 $ from the problem. Equivalently we can set $ \theta_1 =\theta_2=0$.
 
\begin{figure}[]
\centering
\includegraphics[width=85mm]{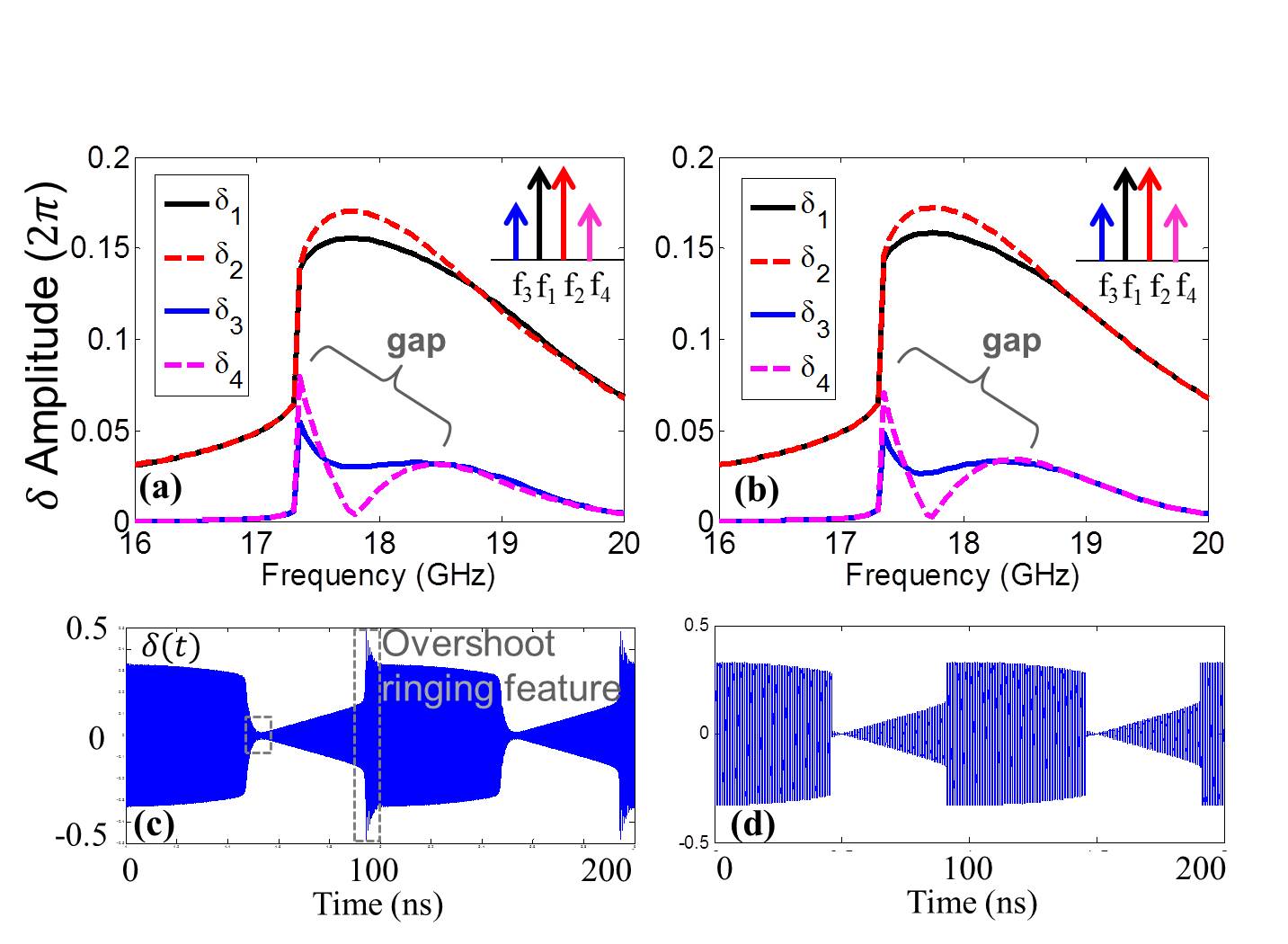}
\caption{The lower and higher main tone output amplitudes $\delta_{{1}}$ and $\delta_{{2}}$, and third order tones $\delta_{{3}}$ and $\delta_{{4}}$ for a single rf-SQUID meta-atom at -65 dBm calculated with (a) numerical simulation and (b) analytical model. Plots of $\delta(t)$ over a beat period at 17.35 GHz and -65 dBm calculated by (c) numerical simulation and (d) steady-state analytical model. The dashed boxes in (c) point out the overshooting ringing features in numerical simulation. The spacing between the two input tones is 10 MHz, the temperature is 4.6 K, and the applied dc flux is set to zero.}
\label{gap_-65dBm}
\end{figure}

We first consider the case of equal amplitude tones (set $\phi_{rf,1}=\phi_{rf,2}=\phi_{s}$ to be the amplitude) and set $\theta_1=\theta_2=0$. We then solve Eq. (\ref{delta_IMD}) for $\delta (\tau)$ using the previously described circuit parameters. Under all circumstances explored here $\delta(\tau)$ is observed to be sinusoidal to a good approximation.  Figure \ref{gap_-65dBm} (c) is an example of the solution to $\delta (\tau)$ at an input power of $-65$ dBm, with tone frequencies $f_1$ and $f_2$ centered around $f=17.35$ GHz and separated by of $\Delta f=10$ MHz. The dense blue curves are the fast carrier oscillations and the vertical extreme of the blue represents the slowly varying envelope. More precisely, $\delta (\tau)$ can be represented as in Eq. (4). In this example, the envelope varies on a time scale 3 orders of magnitude longer than the carrier period. One beat period of the envelope is shown in Fig. \ref{gap_-65dBm} (c) .

Further to investigate IM, we extract the amplitude and phase of $\delta_{{i}}$ for frequency component $f_i$ via Fourier transform of $\delta (\tau)$. Since magnetic flux is related to $\delta$ through $\delta=2\pi(\Phi_{tot}/\Phi_0)$, we can extract the generated third order IM magnetic flux $\Phi_{3,4}$. The IM flux translates into an IM magnetic field inside the SQUID loop of area $A$, i.e., $B_{3,4}=\Phi_{3,4}/A$. The excited IM magnetic field transmits through the rectangular waveguide and generates the third order IM powers at the detector. The SQUID is inductively coupled to the waveguide via a coupling coefficient $g$  \cite{Doyle2008}, so only part of the IM power couples to the waveguide mode. The final simulated output IM power is adjusted by varying $g$ ($g\approx 0.015$ for the single SQUID meta-atom), and plotted as a function of center frequency and tone power in Fig. \ref{IMD_asym} (b) for the upper third order IM tone $P_{f_4}$, with a cut through -65 dBm plotting both lower and upper third order IM powers ($P_{f_3}$ and $P_{f_4}$) in Fig. \ref{IMD_asym} (d). The cut through the simulated IM power displays a similar sharp onset and gap feature as observed in the experiment, as well as the prominent asymmetry between the two IM tones.


Since $\delta_{{i}}$ is a surrogate for the output tone power $P_{f_i}$ ($\delta_i\sim\sqrt{P_{f_i}}$) and a direct solution of the nonlinear equation, we use this quantity to analyze the degree of IM generation. Figure \ref{gap_-65dBm} (a) shows amplitudes of $\delta_{{1}}$ to $\delta_{{4}}$ as a function of tone center frequency at an input power of -65 dBm, which shows the same asymmetric gap feature. The upper third order IM output $\delta_{{4}}$ reduces to nearly zero inside the gap. We plot $\delta(t)$ during one beat period of the input rf signal at the onset center frequency (17.35 GHz) of the abrupt IM generation peak in Fig. 3 (c). The $\delta(t)$ envelope stays at a higher amplitude in the first quarter of the signal beat period, suddenly decreases to a low amplitude, and gradually increases before it jumps to a higher amplitude again. Note that each abrupt jumps comes with an overshoot feature (labeled as dashed boxes in Fig.\ref{gap_-65dBm} (c)) with a frequency around 1.5 GHz. The overshoot frequency is intermediate to the fast oscillation (17.35 GHz) and the slow modulations (10 MHz).

\begin{figure}[]
\centering
\includegraphics[width=85mm]{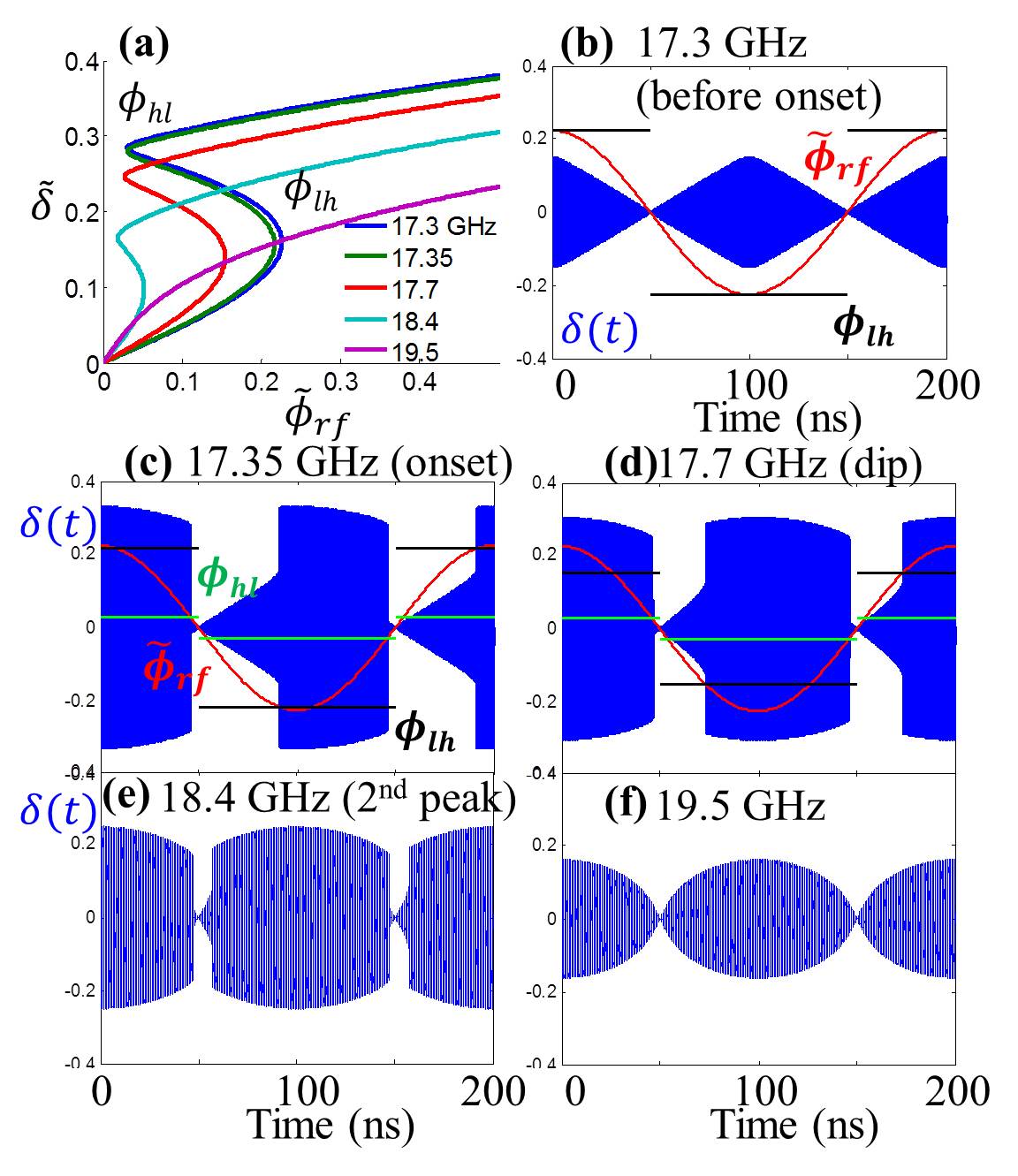}
\caption{Analytical solutions of steady-state model (Eqs. (\ref{delta_ana1})-(\ref{delta_ana3})) at an rf power of -65 dBm which is around the gap feature . (a) The relationship between $\tilde{\delta}$ and $\tilde{\phi}_{rf}$ for five remarkable frequencies. $\phi_{lh}$ denotes the value of rf flux required for transitions of $\tilde{\delta}$ from low to high amplitude solution branch, and $\phi_{hl}$ denotes the rf flux value for the transition from high to low amplitude solution. (b) to (f): Blue curves represent $\delta(t)$ calculated by the analytical model for (a) 17.3 GHz, right before the onset of strong IM generation, (b) 17.35 GHz, at the onset (c) 17.7 GHz, at the gap (d) 18.4 GHz, at the $2^{nd}$ peak, and (e) 19.5 GHz, low IM generation. The red curve is $\tilde{\phi}_{rf}$ as a function of time during a beat period. $\phi_{lh}$ and $\phi_{hl}$ are marked in the figures as black and green lines. All assume $\phi_{dc}=0$.}
\label{ana_solution}
\end{figure}

\subsection{Steady-State Analytical Model}
In this section we develop an analytical model to understand the unique phenomena revealed in the experiment and the numerical solutions of the previous sections. We adopt the observation that the gauge-invariant phase $\delta (\tau)$ and the driving flux can be represented as in Eq. (\ref{phasor}) as a rapidly varying carrier modulated by an envelope. Thus, we insert Eq. (\ref{phasor}) on the right hand side of Eq. (\ref{delta_IMD}). We first look for solutions where the time variation of the envelope is so slow that the temporal derivatives of it can be ignored. This leads (after neglecting harmonics of the drive signal, which will be justified below) to a time dependent gauge-invariant phase:
\begin{equation*}
\delta(\tau)=\bar{\delta}+\tilde{\delta}\sin(\Omega \tau + \theta)
\end{equation*}
where $\Omega=(\omega_1+\omega_2)/(2\omega_{geo})$ and $\bar{\delta}$,  $\tilde{\delta}$ and $\theta$ are taken to be constants that depend parametrically on $\tau$ through the slow variation of $\phi_{rf} (\tau)=\tilde{\phi}_{rf}=\phi_e$. Here $\bar{\delta}$ and $\tilde{\delta}$ denote the dc part and the slowly varying envelope of $\delta$, respectively, $\theta$ is the phase of $\delta$ (which can also vary slowly with time).

For the nonlinear term in Eq. (\ref{delta_IMD}) we have 
$\sin \delta=\sin [\bar{\delta}+\tilde{\delta}\sin(\Omega \tau + \theta)] = \sin\bar{\delta}\cos[\tilde{\delta}\sin(\Omega \tau + \theta)]+cos\bar{\delta}\sin[\tilde{\delta}\sin(\Omega \tau + \theta)]$. 
In principle this term will contain all harmonics of the carrier, $n\Omega$ ($n=0,1,2,...$), and induce harmonics in the gauge-invariant phase $\delta(\tau)$. However, higher harmonics in the gauge-invariant phase are suppressed by the second derivative term  in Eq. (\ref{delta_IMD}) (capacitive current). This is confirmed in our numerical solutions where the amplitudes of higher harmonics (components of frequency $2\Omega$ and $3\Omega$) of $\delta$ are at least 2 orders of magnitude lower than the fundamental frequency component. We note that for the examples considered here the dc phase, $\bar{\delta}$, is zero and consequently only odd harmonics are present. We thus neglect these higher order harmonic terms when we expand $\sin[\tilde{\delta}\sin(\Omega \tau + \theta)]$ and $\cos[\tilde{\delta}\sin(\Omega \tau + \theta)]$. As a result, we obtain  
$\sin \delta \approx \sin \bar{\delta} J_0 (\tilde{\delta})+ 2 \cos \bar{\delta} J_1 (\tilde{\delta}) \sin(\Omega \tau + \theta) $ where $J_0 (\tilde{\delta})$ and $J_1 (\tilde{\delta})$ are Bessel functions. Separating the dc, in-phase, and quadrature components of Eq. (\ref{delta_IMD}), leads to three coupled equations for the three unknowns ($\bar{\delta}$, $\tilde{\delta}$ and $\theta$),

\begin{equation}
\label{delta_ana1}
(1-\Omega^2)\tilde{\delta}+2\beta_{rf}\cos\bar{\delta}J_1(\tilde{\delta})=\tilde{\phi}_{rf}\cos\theta
\end{equation}
\begin{equation}
\label{delta_ana2}
\frac{\Omega}{Q}\tilde{\delta}=-\tilde{\phi}_{rf}\sin\theta
\end{equation}
\begin{equation}
\label{delta_ana3}
\bar{\delta}+\beta_{rf}\sin\bar{\delta}J_0(\tilde{\delta})=\phi_{dc}
\end{equation}

We construct $\delta(t)$ by solving Eqs. (\ref{delta_ana1}) - (\ref{delta_ana3}) to find $\bar{\delta}$, $\tilde{\delta}$, and $\theta$ for a given $\tilde{\phi}_{rf}$ and $\phi_{dc}$. The relationship between $\tilde{\delta}$ and $\tilde{\phi}_{rf}$ at different frequencies $(f_1+f_2)/2$ is plotted in Fig. \ref{ana_solution} (a) for our standard parameter set, $\phi_{dc}=0$, $Q=75$, and $\beta_{rf}=0.98$. The oscillation amplitude $\tilde{\delta}$ as a function of rf flux amplitude $\tilde{\phi}_{rf}$ is symmetric about the origin, so only positive $\tilde{\phi}_{rf}$ is shown. Figure \ref{ana_solution} (a) indicates that $\tilde{\delta}$ can be single-valued or multi-valued depending on the fast-oscillation frequency and the slowly-varying envelope amplitude $\tilde{\phi}_{rf}$. For cases where $\tilde{\delta}$ is multivalued, we let $\phi_{hl}$ and $\phi_{lh}$ denote the lower and upper critical rf flux values (as labeled in Fig. \ref{ana_solution} (a)) between which there are three solutions for the oscillation amplitude, $\tilde{\delta}$. When this occurs ($\phi_{hl}<\tilde{\phi}_{rf}<\phi_{lh}$) the middle solution is always unstable and the largest and the smallest solutions are stable. Thus, if $\tilde{\phi}_{rf}$ is in the bistable regime, and $\tilde{\delta}$ is on the lower (higher) stable branch, then, as $\tilde{\phi}_{rf}$ is slowly increased (decreased) through $\phi_{lh}$ ($\phi_{hl}$), the solution for $\tilde{\delta}$ will experience a jump transition from the lower (higher) stable branch to the higher (lower) stable branch.



For two equal amplitude input tones with a fixed center frequency and a fixed tone power, 
$\tilde{\phi}_{rf}$ is a sinusoidal function with a peak value of $2\phi_{s}$, and a frequency of $\Delta \Omega/2$, i.e. $\tilde{\phi}_{rf}=2\phi_{s}\cos\Delta\Omega\tau/2$. 

Figures \ref{ana_solution} (b) - (f) show the evolution of $\delta(t)$ at different center frequencies (blue), as well as the relationship between the envelopes of the rf flux  $\tilde{\phi}_{rf}$ (red curves), the transition rf flux values $\phi_{lh}$ (black horizontal lines) and $\phi_{hl}$ (green horizontal lines) for positive and negative $\tilde{\phi}_{rf}$ values during a beat period ($\omega_{geo}\Delta \Omega/2\pi=10$ MHz). For tone center frequencies below 17.3 GHz, although $\tilde{\delta}$ is bistable, the envelope of rf flux $\tilde{\phi}_{rf}$ is always below $\phi_{lh}$, so $\tilde{\delta}$ remains on the low amplitude branch during a beat period. Above 18.6 GHz, $\tilde{\delta}$ as a function of $\tilde{\phi}_{rf}$ becomes single valued. Both cases give rise to low IM generation.

Between 17.3 GHz to 18.6 GHz, however, the peak value of $\tilde{\phi}_{rf}$ exceeds the upper bi-stable transition rf flux amplitude $\phi_{lh}$, while the minimum value of $\tilde{\phi}_{rf}$ is below $\phi_{hl}$, so there are four discontinuous jumps in $\delta(t)$ during a beat period. Changing the center frequency from 17.35 GHz to 17.7 GHz makes the  solutions for $\tilde{\delta}$ stay on the high-amplitude branch longer (Fig. \ref{ana_solution} (d)). This is because $\phi_{lh}$ is smaller for higher frequencies (as seen in Fig. \ref{ana_solution} (a)), so it is easier for $\tilde{\phi}_{rf}$ to pass the low-to-high transition. The sudden asymmetric state jumps during a beat period generates rich IM products.

We extract the IM components of $\delta$ by Fourier transform as discussed for the numerical simulation, and extract the amplitude of two main tones and two third order IM tones of $\delta$, plotted in Fig. \ref{gap_-65dBm} (b). The analytically calculated amplitudes of IM tones are almost the same as those in the full numerical simulation. However, comparison of time dependent gauge-invariant phase $\delta(t)$ between the full numerical calculation and the analytical calculation in Fig. \ref{gap_-65dBm} (c) and (d) indicates that the dynamical ringing appears around the state jumps in the full-nonlinear numerical calculation but is not present in the steady-state solutions to Eqs. (\ref{delta_ana1}) to (\ref{delta_ana3}). These will be investigated subsequently.

\subsection{Dynamical Model}

The ringing behavior of $\delta (t)$ during state jumps indicates that the system requires time to transition from one stable state to another. We study this process using a dynamical model for the complex amplitude of the phase $\hat{\delta}$, where ${\delta} (\tau)=\bar{\delta}+Re[\hat{\delta}(\tau) e^{i\Omega\tau-i\pi/2}]$.

For two equal amplitude input tones, the envelope of the rf flux  $\hat{\phi}_{rf}=\phi_e=2\phi_{s}\cos(\Delta\Omega\tau/2)$ is real. In this case, $\sin\delta$ is expanded as $\sin\bar{\delta}J_0(|\hat{\delta}|)+2\cos\bar{\delta}{J_1(|\hat{\delta}|)}Re(\hat{\delta}e^{i\Omega\tau-i\pi/2})/{|\hat{\delta}|}$ with negligible higher order terms assuming that the higher harmonics of $\delta$ are much smaller than the base frequency component. In deriving an equation for the envelope, we adopt the approximations that $Q>>1$ and that $\hat{\delta} (\tau)$ changes slowly, $|\Omega \hat{\delta}| >> |d\hat{\delta}/d\tau|$. Thus in Eq. (\ref{delta_IMD}) we replace ${d}/{(Q d\tau)}$ with $i\Omega/Q$, and ${d^2}/{d\tau^2}$ with $-\Omega^2+2i\Omega{d}/{d\tau}$. This yields a first-order nonlinear equation for the phasor $\hat{\delta}$ and a transcendental equation for the steady part of $\delta(t)$,

\begin{equation}
\label{delta_dyn1}
i\Omega[2\frac{d}{d\tau}+\frac{1}{Q}]\hat{\delta}+[1-\Omega^2+\beta_{rf}\cos\bar{\delta}\frac{2J_1(|\hat{\delta}|)}{|\hat{\delta}|}]\hat{\delta}=\hat{\phi}_{rf}
\end{equation}

\begin{equation}
\label{delta_dyn2}
\bar{\delta}+\beta_{rf}\sin\bar{\delta}J_0(|\hat{\delta}|)=\phi_{dc}
\end{equation}

To analyze the dynamics, we express $\hat{\delta}$ as an in-phase part and a quadrature part, $i.e.$  $\hat{\delta} = \delta_R + i\delta_I$, and write the real and imaginary parts of Eq. (\ref{delta_dyn1}). We note that in the absence of losses ($Q\rightarrow\infty$) one can construct a Hamiltonian function for the nonlinear system. Including losses we have
\begin{equation*}
\tag{10a}
\frac{d}{d\tau}\delta_R=-\frac{1}{2Q}\delta_R-\frac{\partial}{\partial\delta_I}H(|\hat{\delta}|)\\
\end{equation*}
\begin{equation}
\tag{10b}
\label{delta_H}
\frac{d}{d\tau}\delta_I=-\frac{1}{2Q}\delta_I+\frac{\partial}{\partial\delta_R}H(|\hat{\delta}|)
\end{equation}
where 
\begin{equation*}
H=\frac{1}{4\Omega}[(1-\Omega^2){|\hat{\delta}|^2}]-2\beta_{rf}\cos\bar{\delta}J_0(|\hat{\delta}|)-\delta_R\hat{\phi}_{rf}
\end{equation*}
is the Hamiltonian. Equilibrium states of the system Eq. (\ref{delta_H}) are the same as those described by Eqs. (\ref{delta_ana1}) - (\ref{delta_ana3}). However, we note that the Q-value for our system is quite large, $Q\approx75$. As a result we look for equilibria of the lossless system, $Q\rightarrow\infty$, which are located in the $\delta_R-\delta_I$ plane at the stationary values of the Hamiltonian, $\partial H/ \partial \delta_R=\partial H/ \partial \delta_I=0$. Equilibria will be stable if they are at maximal or minimal points of $H$ when $({\partial^2H}/{\partial\delta_I^2})({\partial^2H}/{\partial\delta_R^2})>0$. Note that the Hamiltonian is symmetric about $\delta_I=0$.

\begin{figure}[]
\centering
\includegraphics[width=85mm]{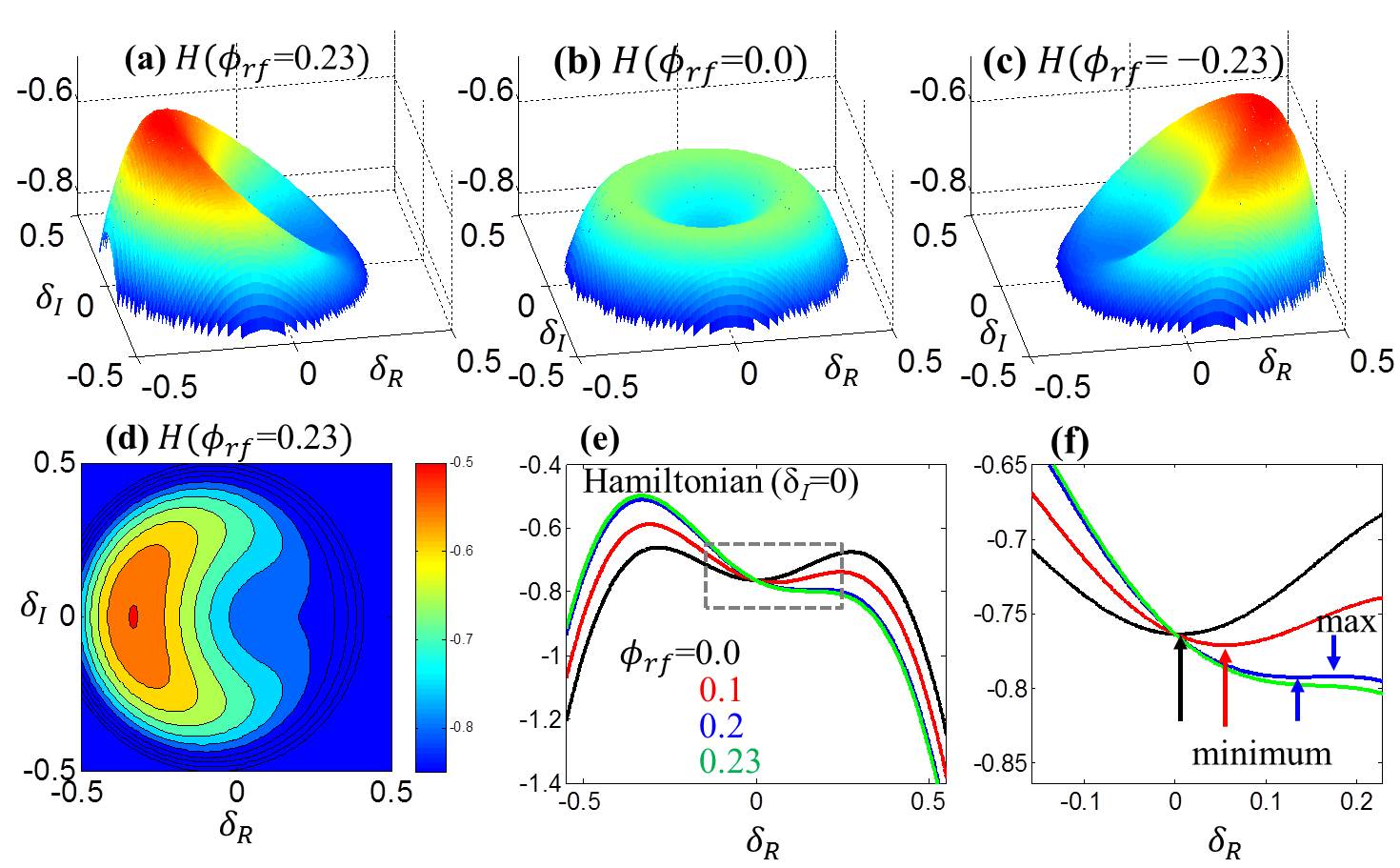}
\caption{The calculated Hamiltonian of a single rf SQUID as a function of $\delta_R$ and $\delta_I$ for rf flux amplitudes of (a) 0.23, (b) 0.0, and (c) -0.23. (d) The colormap of the calculated Hamiltonian as a function of $\delta_R$ and $\delta_I$ for rf flux amplitude of 0.23, with contours from $-1$ to $-0.5$ with a step of $0.05$. (e) The calculated Hamiltonian as a function of $\delta_R$ when $\delta_I=0$ with different values of rf flux. (f) A zoom-in plot of the dashed box in (e). The transition rf flux value to bistability is around 0.22. All assume a center frequency of 17.35 GHz.}
\label{H_delta}
\end{figure}

In Figs. \ref{H_delta} (a) - (c) we plot the Hamiltonian as a function of $\delta_R$ and $\delta_I$ at a center frequency of 17.35 GHz and -65 dBm tone power, when the rf flux amplitude $\hat{\phi}_{rf}$ is at its peak ($0.23$), zero ($0.0$), and negative maximum ($-0.23$) during a beat period. 
Figure \ref{H_delta} (e) shows a cut through the $\delta_I=0$ plane, plotting $H$ as a function of $\delta_R$ at various rf flux values. In Fig. \ref{H_delta} (f) a blow-up of the dashed region is shown that traces the minimum and maximum of $H$ as the rf flux envelope evolves with time. Note that the state transition occurs at an rf flux amplitude of 0.22 for this frequency. When rf flux is zero, the Hamiltonian $H$ is symmetric around the origin, and has a local minimum (stable point) centered at the origin. As the rf flux increases, the $H(\delta_R)$ curve tilts so that the peak located in the positive region of $\delta_R$ decreases and moves towards the origin; gradually meeting the dip which moves away from the origin along the $\delta_R$ axis. At the same time another peak rises up. As the rf flux value reaches 0.23, the lower peak and the dip between the two peaks disappear. The system then has to transition to another stable state located at the higher peak in the negative $\delta_R$ region. At an rf flux of $-0.23$, $H$ tilts to the other side (Fig. \ref{H_delta} (c)). 

Because of the high value of $Q$, the system's transition trajectory from one stable state to another follows the constant contour lines of the Hamiltonian surfaces in a spiral manner. Figure \ref{H_delta} (d) shows the contour lines (from $-1$ to $-0.5$ with a step of $0.05$) on top of the Hamiltonian colormap at $\hat{\phi}_{rf}=0.23$.

\begin{figure}[]
\centering
\includegraphics[width=85mm]{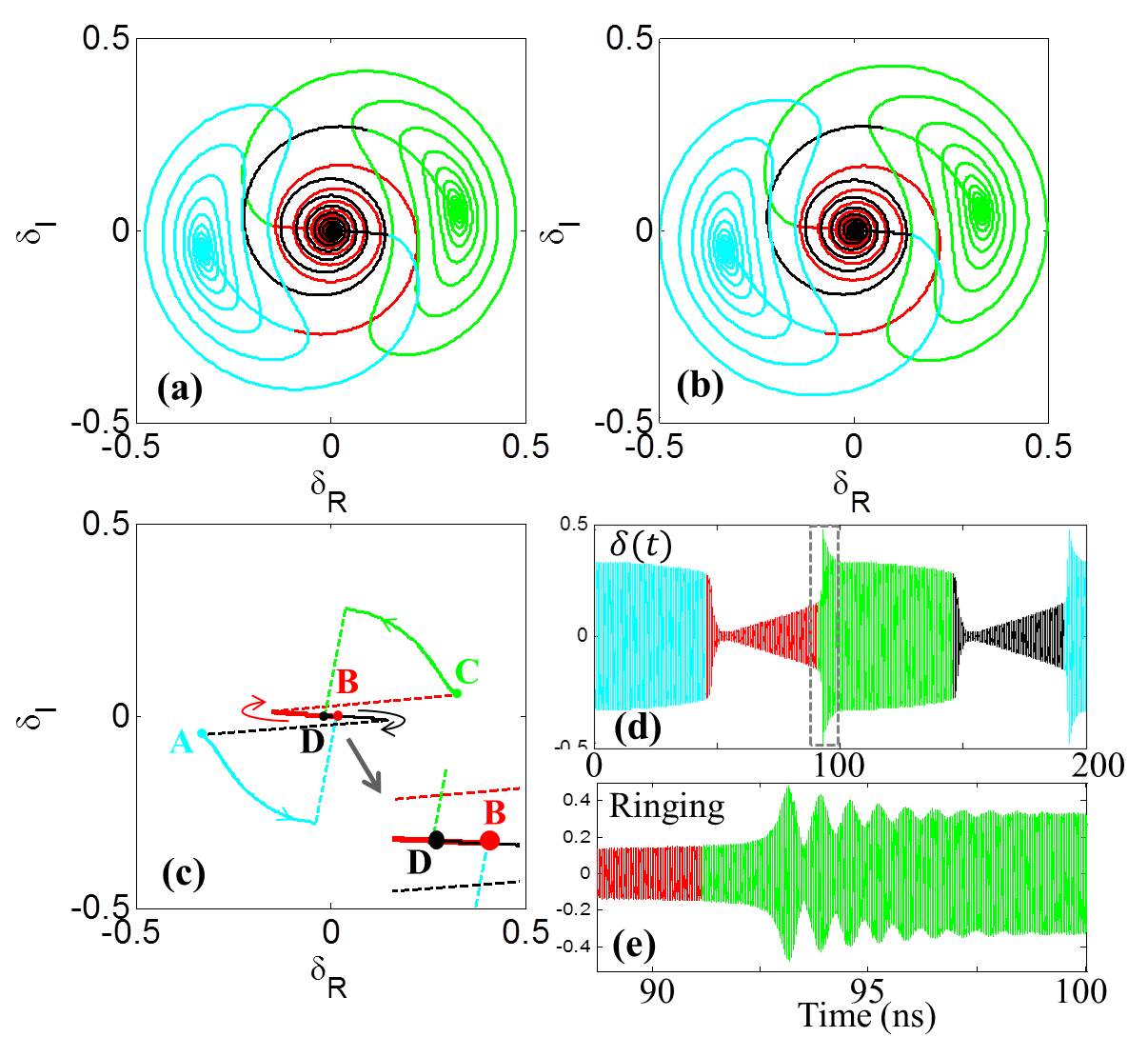}
\caption{The time elapsed trajectories for $\hat{\delta}$(t) for one beat period calculated by (a) the dynamical model, (b) the numerical simulation, and (c) the steady-state model. The inset of (c) zooms in on the trajectory around the origin by five times. (d) shows $\delta (t)$ calculated from the dynamical model, and (e) is a zoom-in of the dashed box in (d) showing the ringing behavior.}
\label{dyn_solution}
\end{figure}

We can find the trajectory of $\hat{\delta} (t)$ for $\phi_{dc}=0$ by solving Eq. (\ref{delta_dyn1}) to obtain $\delta_R$ and $\delta_I$ during a beat period as $\hat{\phi}_{rf}$ changes. Again, we look at the solutions for a center frequency of 17.35 GHz at -65 dBm input tone power. The time trajectory of the phase envelope $\hat{\delta}$ in the ${\delta_R}$-${\delta_I}$ plane during the beat period as calculated by the dynamical model is shown in Fig. \ref{dyn_solution} (a). Compare this with Figs. \ref{dyn_solution} (b) and \ref{dyn_solution} (c) which present the $\hat{\delta}$ trajectories extracted from $\delta(t)$ in the full nonlinear numerical calculation and the steady-state model, respectively. Figure \ref{dyn_solution} (a) and (b) are almost identical to each other, serving to validate the dynamical model. In the trajectory plots Fig. \ref{dyn_solution} (a) and (b)  we see four colored in-spiraling orbits centered around four corresponding dense regions (red and black dense regions are close to each other near the origin); the dense regions denote the steady-state solutions right after a state jump. We can clearly see these four states in the steady-state trajectory (Fig. \ref{dyn_solution} (c)) labeled as $A$, $B$, $C$ and $D$. The blue dense region in Fig. \ref{dyn_solution} (a) and (b) is the solution at the beginning of a beat period, corresponding to state $A$. As the rf flux amplitude during a beat period reduces below $\phi_{hl}$, the high-amplitude state has to jump to state $B$ (red). For the steady-state solution (Fig. \ref{dyn_solution} (c)), the system oscillates in the high-amplitude branch following the blue curve, then directly jumps to state B (red dot). In numerical simulation of Eq. (\ref{delta_IMD}) and the dynamical model Eqs. (10a) and (10b) though, the system goes through several orbits before settling down at the low-amplitude stable state $B$ (red dense region) near the origin in the $\delta$-plane. It follows from Eqs. (10a) and (10b) that the area in phase enclosed by the orbit decreases exponentially at a rate $2/Q$ during approach to the equilibrium point. The boundary between the two colors denotes the time when the system starts to jump to another state. 

The in-spiraling orbits during a transition are predicted by the Hamiltonian analysis. The shape of the trajectory before jumping to state $A$ matches the contour lines in Fig. \ref{H_delta} (d), except that the trajectory is not symmetric about $\delta_I$ axis due to the losses (parameterized by $Q$) which is not included in the Hamiltonian. The number of trajectory orbits during the transition illustrates the relaxation time of a state jump. The relaxation time also depends on the losses.

Figure \ref{dyn_solution} (d) displays the $\delta (t)$ calculated by the dynamical model; Fig. \ref{dyn_solution} (e) is a zoom-in for the selected region near a state jump. The colors match the colored curves in the trajectory plots Fig. \ref{dyn_solution} (a) to (c).  There are very clear ringing features during a jump, which is a reflection of damped spiral orbits. The ringing feature oscillates at a frequency of around 1.5 GHz, and can cause sidebands in the IM spectrum. 

\section{Discussion}
Three models for IM generation in rf-SQUIDs have been discussed. The solutions to the full numerical nonlinear model contain the most complete information for the response of rf-SQUIDs to two-tone excitation, yet gives little insight into the underlying physics. The steady-state analytical model greatly simplifies the $2^{nd}$ order nonlinear differential equation to three coupled algebraic equations, and sheds light on the origin of the unique IM features - the state jumps during a beat period cause an abrupt increase in IM products. While it predicts the same level of IM generation as calculated by numerical simulation (Fig. \ref{gap_-65dBm}), the steady-state model lacks the dynamics accompanying each state jump, which can be understood using the nonlinear dynamical model. This model reduces the full nonlinear equation to a complex first order differential equation, and allows for construction of a Hamiltonian for the SQUID. The topology of the Hamiltonian surfaces evolves continuously as the envelope of the drive signal changes. The topology determines the form of the trajectories, $\hat{\delta} (t)$, to be spirals during transitions as the SQUID switches from one stable state to another, resulting in ringing features in $\delta(t)$.

The models all include dc flux as a variable that affects the response of the SQUID. In this paper we focus on the zero dc flux case. Varying the dc flux value would modify the relationship between the envelope of $\delta$ and the envelope of $\phi_{rf}$ (zero flux case shown in Fig. \ref{ana_solution} (a)), but would preserve bistability and thus the discontinuous jumps during a beat period. In the future we plan to explore the effect of non-zero dc flux on IM generation.

We also note that utilizing two equal-amplitude tone inputs always results in the rf flux envelope passing through zero during the beat period. Thus the IM products of the SQUID are independent of the system's history, even in the bistable regime. As long as the rf flux envelope peak (determined by tone power) exceeds the transition point $\phi_{lh}$, the SQUID will experience four discontinuous jumps during a beat period. However, if the two tones have different amplitudes, so that the minimum value of the envelope is higher than $\phi_{hl}$, the amplitude of the phase envelope depends on the direction of tone power sweep. In an upward sweep the phase amplitude $\hat{\delta}$ resides in the low-amplitude branch during the whole beat period until the tone power increases to the point that the rf flux envelope peak exceeds $\phi_{lh}$; $\delta$  will then keep oscillating in the high-amplitude branch during a beat period. In a downward tone power scan though, $\delta$ would modulate with the beating rf flux in the high-amplitude branch until the peak drops below $\phi_{hl}$. The IM amplification experiment of an $11\times11$ SQUID array metamaterial, where the power amplitude of one tone is always 20 dB higher than the other, shows significantly more hysteresis in rf power scanning than the equal-amplitude IM case. The lack of discontinuous jumps during a beat period in the hysteretic IM amplification process brings in new phenomena worth investigating in the future.

\section{Conclusion}  
We have shown that the rf-SQUID meta-atoms and metamaterials have a rich nonlinear spectrum due to the nonlinearity of the Josephson junctions. Experiment, numerical simulation, and analytic models all show a sharp onset, followed by a dip, in the third order IM output. Rf-SQUID array metamaterials display behaviors that are similar to those of single rf-SQUID meta-atoms. The sharp onset of IM generation comes from a series of asymmetric jumps between two stable states of the rf SQUID as the drive amplitude modulates during a beat period of the input signal. Each state jump creates a transient response appearing as ringing in the time domain. The time evolution of the junction gauge-invariant phase $\delta (t)$  can be explained by a dynamical model employing a Hamiltonian analysis with damping. Our analytical models can potentially be used to design SQUID metamaterials to generate either very high or very low IM products in response to multi-tone excitation. In addition, these models can also be applied to design other nonlinear systems employing Josephson junctions, such as the Josephson parametric amplifiers. 

\begin{acknowledgments}
This work is supported by the NSF-GOALI and OISE programs through grant $\#$ECCS-1158644, and the Center for Nanophysics and Advanced Materials (CNAM). We thank Oleg Mukhanov, Masoud Radparvar, Georgy Prokopenko, Jen-Hao Yeh and Tamin Tai for experimental guidance and helpful suggestions, Hypres Inc. for fabricating the samples, and Alexey Ustinov, Philipp Jung, Susanne Butz for helpful discussions. We also thank H. J. Paik and M. V. Moody for use of the pulsed tube refrigerator.

\end{acknowledgments}

\bibliography{IMD}

\begin{thebibliography}{72}%
\makeatletter
\providecommand \@ifxundefined [1]{%
 \@ifx{#1\undefined}
}%
\providecommand \@ifnum [1]{%
 \ifnum #1\expandafter \@firstoftwo
 \else \expandafter \@secondoftwo
 \fi
}%
\providecommand \@ifx [1]{%
 \ifx #1\expandafter \@firstoftwo
 \else \expandafter \@secondoftwo
 \fi
}%
\providecommand \natexlab [1]{#1}%
\providecommand \enquote  [1]{``#1''}%
\providecommand \bibnamefont  [1]{#1}%
\providecommand \bibfnamefont [1]{#1}%
\providecommand \citenamefont [1]{#1}%
\providecommand \href@noop [0]{\@secondoftwo}%
\providecommand \href [0]{\begingroup \@sanitize@url \@href}%
\providecommand \@href[1]{\@@startlink{#1}\@@href}%
\providecommand \@@href[1]{\endgroup#1\@@endlink}%
\providecommand \@sanitize@url [0]{\catcode `\\12\catcode `\$12\catcode
  `\&12\catcode `\#12\catcode `\^12\catcode `\_12\catcode `\%12\relax}%
\providecommand \@@startlink[1]{}%
\providecommand \@@endlink[0]{}%
\providecommand \url  [0]{\begingroup\@sanitize@url \@url }%
\providecommand \@url [1]{\endgroup\@href {#1}{\urlprefix }}%
\providecommand \urlprefix  [0]{URL }%
\providecommand \Eprint [0]{\href }%
\providecommand \doibase [0]{http://dx.doi.org/}%
\providecommand \selectlanguage [0]{\@gobble}%
\providecommand \bibinfo  [0]{\@secondoftwo}%
\providecommand \bibfield  [0]{\@secondoftwo}%
\providecommand \translation [1]{[#1]}%
\providecommand \BibitemOpen [0]{}%
\providecommand \bibitemStop [0]{}%
\providecommand \bibitemNoStop [0]{.\EOS\space}%
\providecommand \EOS [0]{\spacefactor3000\relax}%
\providecommand \BibitemShut  [1]{\csname bibitem#1\endcsname}%
\let\auto@bib@innerbib\@empty
\bibitem [{\citenamefont {Vijay}\ \emph {et~al.}(2009)\citenamefont {Vijay},
  \citenamefont {Devoret},\ and\ \citenamefont {Siddiqi}}]{Vijay2009}%
  \BibitemOpen
  \bibfield  {author} {\bibinfo {author} {\bibfnamefont {R.}~\bibnamefont
  {Vijay}}, \bibinfo {author} {\bibfnamefont {M.~H.}\ \bibnamefont {Devoret}},
  \ and\ \bibinfo {author} {\bibfnamefont {I.}~\bibnamefont {Siddiqi}},\
  }\bibfield  {title} {\enquote {\bibinfo {title} {Invited review article: The
  {J}osephson bifurcation amplifier},}\ }\href {\doibase
  http://dx.doi.org/10.1063/1.3224703} {\bibfield  {journal} {\bibinfo
  {journal} {Review of Scientific Instruments}\ }\textbf {\bibinfo {volume}
  {80}},\ \bibinfo {eid} {111101} (\bibinfo {year} {2009})}\BibitemShut
  {NoStop}%
\bibitem [{\citenamefont {Byeong Ho~Eom}\ and\ \citenamefont
  {Zmuidzinas}(2012)}]{Eom2012}%
  \BibitemOpen
  \bibfield  {author} {\bibinfo {author} {\bibfnamefont {Henry G.~LeDuc}\
  \bibnamefont {Byeong Ho~Eom}, \bibfnamefont {Peter K.~Day}}\ and\ \bibinfo
  {author} {\bibfnamefont {Jonas}\ \bibnamefont {Zmuidzinas}},\ }\bibfield
  {title} {\enquote {\bibinfo {title} {A wideband, low-noise superconducting
  amplifier with high dynamic range},}\ }\href {\doibase
  http://dx.doi.org/10.1038/nphys2356} {\bibfield  {journal} {\bibinfo
  {journal} {Nat. Phys.}\ }\textbf {\bibinfo {volume} {8}},\ \bibinfo {pages}
  {623 -- 627} (\bibinfo {year} {2012})}\BibitemShut {NoStop}%
\bibitem [{\citenamefont {Yaakobi}\ \emph {et~al.}(2013)\citenamefont
  {Yaakobi}, \citenamefont {Friedland}, \citenamefont {Macklin},\ and\
  \citenamefont {Siddiqi}}]{Siddiqi2013}%
  \BibitemOpen
  \bibfield  {author} {\bibinfo {author} {\bibfnamefont {O.}~\bibnamefont
  {Yaakobi}}, \bibinfo {author} {\bibfnamefont {L.}~\bibnamefont {Friedland}},
  \bibinfo {author} {\bibfnamefont {C.}~\bibnamefont {Macklin}}, \ and\
  \bibinfo {author} {\bibfnamefont {I.}~\bibnamefont {Siddiqi}},\ }\bibfield
  {title} {\enquote {\bibinfo {title} {Parametric amplification in {J}osephson
  junction embedded transmission lines},}\ }\href {\doibase
  10.1103/PhysRevB.87.144301} {\bibfield  {journal} {\bibinfo  {journal} {Phys.
  Rev. B}\ }\textbf {\bibinfo {volume} {87}},\ \bibinfo {pages} {144301}
  (\bibinfo {year} {2013})}\BibitemShut {NoStop}%
\bibitem [{\citenamefont {Mateu}\ \emph {et~al.}(2007)\citenamefont {Mateu},
  \citenamefont {Booth}, \citenamefont {Collado},\ and\ \citenamefont
  {O'Callaghan}}]{Mateu2007}%
  \BibitemOpen
  \bibfield  {author} {\bibinfo {author} {\bibfnamefont {J.}~\bibnamefont
  {Mateu}}, \bibinfo {author} {\bibfnamefont {J.~C.}\ \bibnamefont {Booth}},
  \bibinfo {author} {\bibfnamefont {C.}~\bibnamefont {Collado}}, \ and\
  \bibinfo {author} {\bibfnamefont {J.~M.}\ \bibnamefont {O'Callaghan}},\
  }\bibfield  {title} {\enquote {\bibinfo {title} {Intermodulation distortion
  in coupled-resonator filters with nonuniformly distributed nonlinear
  properties - use in {HTS} {IMD} compensation},}\ }\href {\doibase
  10.1109/TMTT.2007.892802} {\bibfield  {journal} {\bibinfo  {journal} {IEEE
  Transactions on Microwave Theory and Techniques}\ }\textbf {\bibinfo {volume}
  {55}},\ \bibinfo {pages} {616--624} (\bibinfo {year} {2007})}\BibitemShut
  {NoStop}%
\bibitem [{\citenamefont {Lapine}\ \emph {et~al.}(2014)\citenamefont {Lapine},
  \citenamefont {Shadrivov},\ and\ \citenamefont {Kivshar}}]{Lapine2014}%
  \BibitemOpen
  \bibfield  {author} {\bibinfo {author} {\bibfnamefont {Mikhail}\ \bibnamefont
  {Lapine}}, \bibinfo {author} {\bibfnamefont {Ilya~V.}\ \bibnamefont
  {Shadrivov}}, \ and\ \bibinfo {author} {\bibfnamefont {Yuri~S.}\ \bibnamefont
  {Kivshar}},\ }\bibfield  {title} {\enquote {\bibinfo {title}
  {\textit{Colloquium} : Nonlinear metamaterials},}\ }\href {\doibase
  10.1103/RevModPhys.86.1093} {\bibfield  {journal} {\bibinfo  {journal} {Rev.
  Mod. Phys.}\ }\textbf {\bibinfo {volume} {86}},\ \bibinfo {pages}
  {1093--1123} (\bibinfo {year} {2014})}\BibitemShut {NoStop}%
\bibitem [{\citenamefont {Shadrivov}\ \emph {et~al.}(2008)\citenamefont
  {Shadrivov}, \citenamefont {Kozyrev}, \citenamefont {van~der Weide},\ and\
  \citenamefont {Kivshar}}]{Shadrivov2008}%
  \BibitemOpen
  \bibfield  {author} {\bibinfo {author} {\bibfnamefont {Ilya~V.}\ \bibnamefont
  {Shadrivov}}, \bibinfo {author} {\bibfnamefont {Alexander~B.}\ \bibnamefont
  {Kozyrev}}, \bibinfo {author} {\bibfnamefont {Daniel~W.}\ \bibnamefont
  {van~der Weide}}, \ and\ \bibinfo {author} {\bibfnamefont {Yuri~S.}\
  \bibnamefont {Kivshar}},\ }\bibfield  {title} {\enquote {\bibinfo {title}
  {Tunable transmission and harmonic generation in nonlinear metamaterials},}\
  }\href
  {http://scitation.aip.org/content/aip/journal/apl/93/16/10.1063/1.2999634}
  {\bibfield  {journal} {\bibinfo  {journal} {Applied Physics Letters}\
  }\textbf {\bibinfo {volume} {93}},\ \bibinfo {eid} {161903} (\bibinfo {year}
  {2008})}\BibitemShut {NoStop}%
\bibitem [{\citenamefont {Irwin}\ \emph {et~al.}(2008)\citenamefont {Irwin},
  \citenamefont {Hilton}, \citenamefont {Vale},\ and\ \citenamefont
  {Lehnert}}]{Castellanos2008}%
  \BibitemOpen
  \bibfield  {author} {\bibinfo {author} {\bibfnamefont {K.~D.}\ \bibnamefont
  {Irwin}}, \bibinfo {author} {\bibfnamefont {G.~C.}\ \bibnamefont {Hilton}},
  \bibinfo {author} {\bibfnamefont {L.~R.}\ \bibnamefont {Vale}}, \ and\
  \bibinfo {author} {\bibfnamefont {K.~W.}\ \bibnamefont {Lehnert}},\
  }\bibfield  {title} {\enquote {\bibinfo {title} {Amplification and squeezing
  of quantum noise with a tunable {Josephson} metamaterial},}\ }\href {\doibase
  http://dx.doi.org/10.1038/nphys1090} {\bibfield  {journal} {\bibinfo
  {journal} {Nat. Phys.}\ }\textbf {\bibinfo {volume} {4}},\ \bibinfo {pages}
  {929 -- 931} (\bibinfo {year} {2008})}\BibitemShut {NoStop}%
\bibitem [{\citenamefont {Lee}\ and\ \citenamefont {Seo}(2013)}]{Lee2013}%
  \BibitemOpen
  \bibfield  {author} {\bibinfo {author} {\bibfnamefont {Chongmin}\
  \bibnamefont {Lee}}\ and\ \bibinfo {author} {\bibfnamefont {Chulhun}\
  \bibnamefont {Seo}},\ }\bibfield  {title} {\enquote {\bibinfo {title}
  {Control scheme of harmonics and third-order intermodulation distortion with
  composite right/left-handed structure for linearity enhancement of power
  amplifier module},}\ }\href {\doibase 10.1002/mop.27625} {\bibfield
  {journal} {\bibinfo  {journal} {Microwave and Optical Technology Letters}\
  }\textbf {\bibinfo {volume} {55}},\ \bibinfo {pages} {1497--1500} (\bibinfo
  {year} {2013})}\BibitemShut {NoStop}%
\bibitem [{\citenamefont {Gil}\ \emph {et~al.}(2005)\citenamefont {Gil},
  \citenamefont {Bonache}, \citenamefont {Garcia-Garcia}, \citenamefont
  {Falcone},\ and\ \citenamefont {Martin}}]{Gil2005}%
  \BibitemOpen
  \bibfield  {author} {\bibinfo {author} {\bibfnamefont {I.}~\bibnamefont
  {Gil}}, \bibinfo {author} {\bibfnamefont {J.}~\bibnamefont {Bonache}},
  \bibinfo {author} {\bibfnamefont {J.}~\bibnamefont {Garcia-Garcia}}, \bibinfo
  {author} {\bibfnamefont {F.}~\bibnamefont {Falcone}}, \ and\ \bibinfo
  {author} {\bibfnamefont {F.}~\bibnamefont {Martin}},\ }\bibfield  {title}
  {\enquote {\bibinfo {title} {Metamaterials in microstrip technology for
  filter applications},}\ }\bibfield  {booktitle} {\emph {\bibinfo {booktitle}
  {2005 IEEE Antennas and Propagation Society International Symposium}},\
  }\href {\doibase 10.1109/APS.2005.1551409} {\ \textbf {\bibinfo {volume}
  {1A}},\ \bibinfo {pages} {668--671} (\bibinfo {year} {2005})}\BibitemShut
  {NoStop}%
\bibitem [{\citenamefont {Gil}\ \emph {et~al.}(2008)\citenamefont {Gil},
  \citenamefont {Bonache},\ and\ \citenamefont
  {Mart\'{i}n}}]{Meta_filters_review2008}%
  \BibitemOpen
  \bibfield  {author} {\bibinfo {author} {\bibfnamefont {M.}~\bibnamefont
  {Gil}}, \bibinfo {author} {\bibfnamefont {J.}~\bibnamefont {Bonache}}, \ and\
  \bibinfo {author} {\bibfnamefont {F.}~\bibnamefont {Mart\'{i}n}},\ }\bibfield
   {title} {\enquote {\bibinfo {title} {Metamaterial filters: A review},}\
  }\href {\doibase http://dx.doi.org/10.1016/j.metmat.2008.07.006} {\bibfield
  {journal} {\bibinfo  {journal} {Metamaterials}\ }\textbf {\bibinfo {volume}
  {2}},\ \bibinfo {pages} {186 -- 197} (\bibinfo {year} {2008})}\BibitemShut
  {NoStop}%
\bibitem [{\citenamefont {Watts}\ \emph {et~al.}(2016)\citenamefont {Watts},
  \citenamefont {Nadell}, \citenamefont {Montoya}, \citenamefont {Krishna},\
  and\ \citenamefont {Padilla}}]{Watts2016}%
  \BibitemOpen
  \bibfield  {author} {\bibinfo {author} {\bibfnamefont {Claire~M.}\
  \bibnamefont {Watts}}, \bibinfo {author} {\bibfnamefont {Christian~C.}\
  \bibnamefont {Nadell}}, \bibinfo {author} {\bibfnamefont {John}\ \bibnamefont
  {Montoya}}, \bibinfo {author} {\bibfnamefont {Sanjay}\ \bibnamefont
  {Krishna}}, \ and\ \bibinfo {author} {\bibfnamefont {Willie~J.}\ \bibnamefont
  {Padilla}},\ }\bibfield  {title} {\enquote {\bibinfo {title}
  {Frequency-division-multiplexed single-pixel imaging with metamaterials},}\
  }\href {\doibase 10.1364/OPTICA.3.000133} {\bibfield  {journal} {\bibinfo
  {journal} {Optica}\ }\textbf {\bibinfo {volume} {3}},\ \bibinfo {pages}
  {133--138} (\bibinfo {year} {2016})}\BibitemShut {NoStop}%
\bibitem [{\citenamefont {Lim}\ \emph {et~al.}(2005)\citenamefont {Lim},
  \citenamefont {Caloz},\ and\ \citenamefont {Itoh}}]{Lim2005}%
  \BibitemOpen
  \bibfield  {author} {\bibinfo {author} {\bibfnamefont {Sungjoon}\
  \bibnamefont {Lim}}, \bibinfo {author} {\bibfnamefont {C.}~\bibnamefont
  {Caloz}}, \ and\ \bibinfo {author} {\bibfnamefont {T.}~\bibnamefont {Itoh}},\
  }\bibfield  {title} {\enquote {\bibinfo {title} {Metamaterial-based
  electronically controlled transmission-line structure as a novel leaky-wave
  antenna with tunable radiation angle and beamwidth},}\ }\href {\doibase
  10.1109/TMTT.2004.839927} {\bibfield  {journal} {\bibinfo  {journal} {IEEE
  Transactions on Microwave Theory and Techniques}\ }\textbf {\bibinfo {volume}
  {53}},\ \bibinfo {pages} {161--173} (\bibinfo {year} {2005})}\BibitemShut
  {NoStop}%
\bibitem [{\citenamefont {Ziolkowski}\ and\ \citenamefont
  {Erentok}(2006)}]{Ziolkowski2006}%
  \BibitemOpen
  \bibfield  {author} {\bibinfo {author} {\bibfnamefont {Richard~W.}\
  \bibnamefont {Ziolkowski}}\ and\ \bibinfo {author} {\bibfnamefont {Aycan}\
  \bibnamefont {Erentok}},\ }\bibfield  {title} {\enquote {\bibinfo {title}
  {Metamaterial-based efficient electrically small antennas},}\ }\href
  {\doibase 10.1109/TAP.2006.877179} {\bibfield  {journal} {\bibinfo  {journal}
  {IEEE Transactions on Antennas and Propagation}\ }\textbf {\bibinfo {volume}
  {54}},\ \bibinfo {pages} {2113--2130} (\bibinfo {year} {2006})}\BibitemShut
  {NoStop}%
\bibitem [{\citenamefont {Dong}\ \emph {et~al.}(2012)\citenamefont {Dong},
  \citenamefont {Toyao},\ and\ \citenamefont {Itoh}}]{Dong2012}%
  \BibitemOpen
  \bibfield  {author} {\bibinfo {author} {\bibfnamefont {Y.}~\bibnamefont
  {Dong}}, \bibinfo {author} {\bibfnamefont {H.}~\bibnamefont {Toyao}}, \ and\
  \bibinfo {author} {\bibfnamefont {T.}~\bibnamefont {Itoh}},\ }\bibfield
  {title} {\enquote {\bibinfo {title} {Design and characterization of
  miniaturized patch antennas loaded with complementary split-ring
  resonators},}\ }\href {\doibase 10.1109/TAP.2011.2173120} {\bibfield
  {journal} {\bibinfo  {journal} {IEEE Transactions on Antennas and
  Propagation}\ }\textbf {\bibinfo {volume} {60}},\ \bibinfo {pages} {772--785}
  (\bibinfo {year} {2012})}\BibitemShut {NoStop}%
\bibitem [{\citenamefont {Pedro}\ and\ \citenamefont
  {Carvalho}(2002)}]{pedro2002intermodulation}%
  \BibitemOpen
  \bibfield  {author} {\bibinfo {author} {\bibfnamefont {Jos{\'e}~Carlos}\
  \bibnamefont {Pedro}}\ and\ \bibinfo {author} {\bibfnamefont {Nuno~Borges}\
  \bibnamefont {Carvalho}},\ }\href@noop {} {\emph {\bibinfo {title}
  {Intermodulation distortion in microwave and wireless circuits}}}\ (\bibinfo
  {publisher} {Artech House},\ \bibinfo {year} {2002})\BibitemShut {NoStop}%
\bibitem [{\citenamefont {Shen}(1994)}]{shen1994high}%
  \BibitemOpen
  \bibfield  {author} {\bibinfo {author} {\bibfnamefont {Z.Y.}\ \bibnamefont
  {Shen}},\ }\href {https://books.google.com/books?id=qzBTAAAAMAAJ} {\emph
  {\bibinfo {title} {High-temperature Superconducting Microwave Circuits}}},\
  Artech House Antennas and Propagation Library\ (\bibinfo  {publisher} {Artech
  House},\ \bibinfo {year} {1994})\BibitemShut {NoStop}%
\bibitem [{\citenamefont {Abuelma'atti}(1993)}]{Abuelma'atti1993}%
  \BibitemOpen
  \bibfield  {author} {\bibinfo {author} {\bibfnamefont {Muhammad~Taher}\
  \bibnamefont {Abuelma'atti}},\ }\bibfield  {title} {\enquote {\bibinfo
  {title} {Harmonic and intermodulation performance of {J}osephson
  junctions},}\ }\href {\doibase 10.1007/BF02146258} {\bibfield  {journal}
  {\bibinfo  {journal} {International Journal of Infrared and Millimeter
  Waves}\ }\textbf {\bibinfo {volume} {14}},\ \bibinfo {pages} {1299--1310}
  (\bibinfo {year} {1993})}\BibitemShut {NoStop}%
\bibitem [{\citenamefont {Willemsen}\ \emph {et~al.}(1997)\citenamefont
  {Willemsen}, \citenamefont {Dahm},\ and\ \citenamefont
  {Scalapino}}]{Willemsen1997}%
  \BibitemOpen
  \bibfield  {author} {\bibinfo {author} {\bibfnamefont {Balam~A.}\
  \bibnamefont {Willemsen}}, \bibinfo {author} {\bibfnamefont {T.}~\bibnamefont
  {Dahm}}, \ and\ \bibinfo {author} {\bibfnamefont {D.~J.}\ \bibnamefont
  {Scalapino}},\ }\bibfield  {title} {\enquote {\bibinfo {title} {Microwave
  intermodulation in thin film high{-T$_c$} superconducting microstrip hairpin
  resonators: Experiment and theory},}\ }\href {\doibase
  http://dx.doi.org/10.1063/1.120537} {\bibfield  {journal} {\bibinfo
  {journal} {Applied Physics Letters}\ }\textbf {\bibinfo {volume} {71}},\
  \bibinfo {pages} {3898--3900} (\bibinfo {year} {1997})}\BibitemShut {NoStop}%
\bibitem [{\citenamefont {Dahm}\ and\ \citenamefont
  {Scalapino}(1997{\natexlab{a}})}]{Dahm1997JAP}%
  \BibitemOpen
  \bibfield  {author} {\bibinfo {author} {\bibfnamefont {T.}~\bibnamefont
  {Dahm}}\ and\ \bibinfo {author} {\bibfnamefont {D.~J.}\ \bibnamefont
  {Scalapino}},\ }\bibfield  {title} {\enquote {\bibinfo {title}
  {Intermodulation and quality factor of high{-T$_c$} superconducting
  microstrip structures},}\ }\href {\doibase
  http://dx.doi.org/10.1063/1.365839} {\bibfield  {journal} {\bibinfo
  {journal} {Journal of Applied Physics}\ }\textbf {\bibinfo {volume} {82}},\
  \bibinfo {pages} {464--468} (\bibinfo {year}
  {1997}{\natexlab{a}})}\BibitemShut {NoStop}%
\bibitem [{\citenamefont {McDonald}\ \emph {et~al.}(1998)\citenamefont
  {McDonald}, \citenamefont {Clem},\ and\ \citenamefont
  {Oates}}]{McDonald1998}%
  \BibitemOpen
  \bibfield  {author} {\bibinfo {author} {\bibfnamefont {J.}~\bibnamefont
  {McDonald}}, \bibinfo {author} {\bibfnamefont {J.~R.}\ \bibnamefont {Clem}},
  \ and\ \bibinfo {author} {\bibfnamefont {D.~E.}\ \bibnamefont {Oates}},\
  }\bibfield  {title} {\enquote {\bibinfo {title} {Critical-state model for
  intermodulation distortion in a superconducting microwave resonator},}\
  }\href {\doibase http://dx.doi.org/10.1063/1.367356} {\bibfield  {journal}
  {\bibinfo  {journal} {Journal of Applied Physics}\ }\textbf {\bibinfo
  {volume} {83}},\ \bibinfo {pages} {5307--5312} (\bibinfo {year}
  {1998})}\BibitemShut {NoStop}%
\bibitem [{\citenamefont {Hammond}\ \emph {et~al.}(1998)\citenamefont
  {Hammond}, \citenamefont {Soares}, \citenamefont {Willemsen}, \citenamefont
  {Dahm}, \citenamefont {Scalapino},\ and\ \citenamefont
  {Schrieffer}}]{Hammond1998}%
  \BibitemOpen
  \bibfield  {author} {\bibinfo {author} {\bibfnamefont {R.~B.}\ \bibnamefont
  {Hammond}}, \bibinfo {author} {\bibfnamefont {E.~R.}\ \bibnamefont {Soares}},
  \bibinfo {author} {\bibfnamefont {Balam~A.}\ \bibnamefont {Willemsen}},
  \bibinfo {author} {\bibfnamefont {T.}~\bibnamefont {Dahm}}, \bibinfo {author}
  {\bibfnamefont {D.~J.}\ \bibnamefont {Scalapino}}, \ and\ \bibinfo {author}
  {\bibfnamefont {J.~R.}\ \bibnamefont {Schrieffer}},\ }\bibfield  {title}
  {\enquote {\bibinfo {title} {Intrinsic limits on the {Q} and intermodulation
  of low power high temperature superconducting microstrip resonators},}\
  }\href {\doibase http://dx.doi.org/10.1063/1.368827} {\bibfield  {journal}
  {\bibinfo  {journal} {Journal of Applied Physics}\ }\textbf {\bibinfo
  {volume} {84}},\ \bibinfo {pages} {5662--5667} (\bibinfo {year}
  {1998})}\BibitemShut {NoStop}%
\bibitem [{\citenamefont {Benz}\ \emph {et~al.}(1999)\citenamefont {Benz},
  \citenamefont {Scherer}, \citenamefont {Neuhaus},\ and\ \citenamefont
  {Jutzi}}]{Benz1999}%
  \BibitemOpen
  \bibfield  {author} {\bibinfo {author} {\bibfnamefont {G.}~\bibnamefont
  {Benz}}, \bibinfo {author} {\bibfnamefont {T.~A.}\ \bibnamefont {Scherer}},
  \bibinfo {author} {\bibfnamefont {M.}~\bibnamefont {Neuhaus}}, \ and\
  \bibinfo {author} {\bibfnamefont {W.}~\bibnamefont {Jutzi}},\ }\bibfield
  {title} {\enquote {\bibinfo {title} {Quality factor and intermodulation
  product of superconducting coplanar wave guides with slots in a {DC} magnetic
  field},}\ }\href {\doibase 10.1109/77.783671} {\bibfield  {journal} {\bibinfo
   {journal} {IEEE Transactions on Applied Superconductivity}\ }\textbf
  {\bibinfo {volume} {9}},\ \bibinfo {pages} {3046--3049} (\bibinfo {year}
  {1999})}\BibitemShut {NoStop}%
\bibitem [{\citenamefont {Remillard}\ \emph {et~al.}(2003)\citenamefont
  {Remillard}, \citenamefont {ren Yi},\ and\ \citenamefont
  {Abdelmonem}}]{Remillard2003}%
  \BibitemOpen
  \bibfield  {author} {\bibinfo {author} {\bibfnamefont {S.~K.}\ \bibnamefont
  {Remillard}}, \bibinfo {author} {\bibfnamefont {Huai}\ \bibnamefont {ren
  Yi}}, \ and\ \bibinfo {author} {\bibfnamefont {A.}~\bibnamefont
  {Abdelmonem}},\ }\bibfield  {title} {\enquote {\bibinfo {title} {Three-tone
  intermodulation distortion generated by superconducting bandpass filters},}\
  }\href {\doibase 10.1109/TASC.2003.816205} {\bibfield  {journal} {\bibinfo
  {journal} {IEEE Transactions on Applied Superconductivity}\ }\textbf
  {\bibinfo {volume} {13}},\ \bibinfo {pages} {3797--3802} (\bibinfo {year}
  {2003})}\BibitemShut {NoStop}%
\bibitem [{\citenamefont {Mateu}\ \emph {et~al.}(2003)\citenamefont {Mateu},
  \citenamefont {Collado}, \citenamefont {Mené\'{e}ndez},\ and\ \citenamefont
  {O'Callaghan}}]{Mateu2003}%
  \BibitemOpen
  \bibfield  {author} {\bibinfo {author} {\bibfnamefont {J.}~\bibnamefont
  {Mateu}}, \bibinfo {author} {\bibfnamefont {C.}~\bibnamefont {Collado}},
  \bibinfo {author} {\bibfnamefont {O.}~\bibnamefont {Mené\'{e}ndez}}, \ and\
  \bibinfo {author} {\bibfnamefont {J.~M.}\ \bibnamefont {O'Callaghan}},\
  }\bibfield  {title} {\enquote {\bibinfo {title} {A general approach for the
  calculation of intermodulation distortion in cavities with superconducting
  endplates},}\ }\href {\doibase http://dx.doi.org/10.1063/1.1535742}
  {\bibfield  {journal} {\bibinfo  {journal} {Applied Physics Letters}\
  }\textbf {\bibinfo {volume} {82}},\ \bibinfo {pages} {97--99} (\bibinfo
  {year} {2003})}\BibitemShut {NoStop}%
\bibitem [{\citenamefont {Mateu}\ \emph {et~al.}(2009)\citenamefont {Mateu},
  \citenamefont {Collado}, \citenamefont {Orloff}, \citenamefont {Booth},
  \citenamefont {Rocas}, \citenamefont {Padilla},\ and\ \citenamefont
  {O'Callaghan}}]{Mateu2009}%
  \BibitemOpen
  \bibfield  {author} {\bibinfo {author} {\bibfnamefont {J.}~\bibnamefont
  {Mateu}}, \bibinfo {author} {\bibfnamefont {C.}~\bibnamefont {Collado}},
  \bibinfo {author} {\bibfnamefont {N.}~\bibnamefont {Orloff}}, \bibinfo
  {author} {\bibfnamefont {J.~C.}\ \bibnamefont {Booth}}, \bibinfo {author}
  {\bibfnamefont {E.}~\bibnamefont {Rocas}}, \bibinfo {author} {\bibfnamefont
  {A.}~\bibnamefont {Padilla}}, \ and\ \bibinfo {author} {\bibfnamefont
  {J.~M.}\ \bibnamefont {O'Callaghan}},\ }\bibfield  {title} {\enquote
  {\bibinfo {title} {Third-order intermodulation distortion and harmonic
  generation in mismatched weakly nonlinear transmission lines},}\ }\href
  {\doibase 10.1109/TMTT.2008.2009083} {\bibfield  {journal} {\bibinfo
  {journal} {IEEE Transactions on Microwave Theory and Techniques}\ }\textbf
  {\bibinfo {volume} {57}},\ \bibinfo {pages} {10--18} (\bibinfo {year}
  {2009})}\BibitemShut {NoStop}%
\bibitem [{\citenamefont {Rocas}\ \emph {et~al.}(2011)\citenamefont {Rocas},
  \citenamefont {Collado}, \citenamefont {Orloff}, \citenamefont {Mateu},
  \citenamefont {Padilla}, \citenamefont {O'Callaghan},\ and\ \citenamefont
  {Booth}}]{Rocas2011}%
  \BibitemOpen
  \bibfield  {author} {\bibinfo {author} {\bibfnamefont {E.}~\bibnamefont
  {Rocas}}, \bibinfo {author} {\bibfnamefont {C.}~\bibnamefont {Collado}},
  \bibinfo {author} {\bibfnamefont {N.~D.}\ \bibnamefont {Orloff}}, \bibinfo
  {author} {\bibfnamefont {J.}~\bibnamefont {Mateu}}, \bibinfo {author}
  {\bibfnamefont {A.}~\bibnamefont {Padilla}}, \bibinfo {author} {\bibfnamefont
  {J.~M.}\ \bibnamefont {O'Callaghan}}, \ and\ \bibinfo {author} {\bibfnamefont
  {J.~C.}\ \bibnamefont {Booth}},\ }\bibfield  {title} {\enquote {\bibinfo
  {title} {Passive intermodulation due to self-heating in printed transmission
  lines},}\ }\href {\doibase 10.1109/TMTT.2010.2090356} {\bibfield  {journal}
  {\bibinfo  {journal} {IEEE Transactions on Microwave Theory and Techniques}\
  }\textbf {\bibinfo {volume} {59}},\ \bibinfo {pages} {311--322} (\bibinfo
  {year} {2011})}\BibitemShut {NoStop}%
\bibitem [{\citenamefont {Thol{\'e}n}\ \emph {et~al.}(2014)\citenamefont
  {Thol{\'e}n}, \citenamefont {Erg{\"u}l}, \citenamefont {Schaeffer},\ and\
  \citenamefont {Haviland}}]{Tholén2014}%
  \BibitemOpen
  \bibfield  {author} {\bibinfo {author} {\bibfnamefont {Erik~A.}\ \bibnamefont
  {Thol{\'e}n}}, \bibinfo {author} {\bibfnamefont {Adem}\ \bibnamefont
  {Erg{\"u}l}}, \bibinfo {author} {\bibfnamefont {David}\ \bibnamefont
  {Schaeffer}}, \ and\ \bibinfo {author} {\bibfnamefont {David~B.}\
  \bibnamefont {Haviland}},\ }\bibfield  {title} {\enquote {\bibinfo {title}
  {Gain, noise and intermodulation in a nonlinear superconducting resonator},}\
  }\href {\doibase 10.1140/epjqt5} {\bibfield  {journal} {\bibinfo  {journal}
  {EPJ Quantum Technology}\ }\textbf {\bibinfo {volume} {1}},\ \bibinfo {pages}
  {1--10} (\bibinfo {year} {2014})}\BibitemShut {NoStop}%
\bibitem [{\citenamefont {Oates}\ \emph {et~al.}(2001)\citenamefont {Oates},
  \citenamefont {Xin}, \citenamefont {Dresselhaus},\ and\ \citenamefont
  {Dresselhaus}}]{Oates2001}%
  \BibitemOpen
  \bibfield  {author} {\bibinfo {author} {\bibfnamefont {D.~E.}\ \bibnamefont
  {Oates}}, \bibinfo {author} {\bibfnamefont {Hao}\ \bibnamefont {Xin}},
  \bibinfo {author} {\bibfnamefont {G.}~\bibnamefont {Dresselhaus}}, \ and\
  \bibinfo {author} {\bibfnamefont {M.~S.}\ \bibnamefont {Dresselhaus}},\
  }\bibfield  {title} {\enquote {\bibinfo {title} {Intermodulation distortion
  and {J}osephson vortices in {YBCO} bicrystal grain boundaries},}\ }\href
  {\doibase 10.1109/77.919646} {\bibfield  {journal} {\bibinfo  {journal} {IEEE
  Transactions on Applied Superconductivity}\ }\textbf {\bibinfo {volume}
  {11}},\ \bibinfo {pages} {2804--2807} (\bibinfo {year} {2001})}\BibitemShut
  {NoStop}%
\bibitem [{\citenamefont {Oates}\ \emph {et~al.}(2003)\citenamefont {Oates},
  \citenamefont {Park}, \citenamefont {Hein}, \citenamefont {Hirst},\ and\
  \citenamefont {Humphreys}}]{Oates2003}%
  \BibitemOpen
  \bibfield  {author} {\bibinfo {author} {\bibfnamefont {D.~E.}\ \bibnamefont
  {Oates}}, \bibinfo {author} {\bibfnamefont {S.~H.}\ \bibnamefont {Park}},
  \bibinfo {author} {\bibfnamefont {M.~A.}\ \bibnamefont {Hein}}, \bibinfo
  {author} {\bibfnamefont {P.~J.}\ \bibnamefont {Hirst}}, \ and\ \bibinfo
  {author} {\bibfnamefont {R.~G.}\ \bibnamefont {Humphreys}},\ }\bibfield
  {title} {\enquote {\bibinfo {title} {Intermodulation distortion and
  third-harmonic generation in {YBCO} films of varying oxygen content},}\
  }\href {\doibase 10.1109/TASC.2003.813717} {\bibfield  {journal} {\bibinfo
  {journal} {IEEE Transactions on Applied Superconductivity}\ }\textbf
  {\bibinfo {volume} {13}},\ \bibinfo {pages} {311--314} (\bibinfo {year}
  {2003})}\BibitemShut {NoStop}%
\bibitem [{\citenamefont {Zhuravel}\ \emph {et~al.}(2003)\citenamefont
  {Zhuravel}, \citenamefont {Ustinov}, \citenamefont {Abraimov},\ and\
  \citenamefont {Anlage}}]{Zhuravel2003}%
  \BibitemOpen
  \bibfield  {author} {\bibinfo {author} {\bibfnamefont {A.~P.}\ \bibnamefont
  {Zhuravel}}, \bibinfo {author} {\bibfnamefont {A.~V.}\ \bibnamefont
  {Ustinov}}, \bibinfo {author} {\bibfnamefont {D.}~\bibnamefont {Abraimov}}, \
  and\ \bibinfo {author} {\bibfnamefont {S.~M.}\ \bibnamefont {Anlage}},\
  }\bibfield  {title} {\enquote {\bibinfo {title} {Imaging local sources of
  intermodulation in superconducting microwave devices},}\ }\href {\doibase
  10.1109/TASC.2003.813731} {\bibfield  {journal} {\bibinfo  {journal} {IEEE
  Transactions on Applied Superconductivity}\ }\textbf {\bibinfo {volume}
  {13}},\ \bibinfo {pages} {340--343} (\bibinfo {year} {2003})}\BibitemShut
  {NoStop}%
\bibitem [{\citenamefont {Xin}\ \emph {et~al.}(2002)\citenamefont {Xin},
  \citenamefont {Oates}, \citenamefont {Dresselhaus},\ and\ \citenamefont
  {Dresselhaus}}]{Xin2002}%
  \BibitemOpen
  \bibfield  {author} {\bibinfo {author} {\bibfnamefont {H.}~\bibnamefont
  {Xin}}, \bibinfo {author} {\bibfnamefont {D.~E.}\ \bibnamefont {Oates}},
  \bibinfo {author} {\bibfnamefont {G.}~\bibnamefont {Dresselhaus}}, \ and\
  \bibinfo {author} {\bibfnamefont {M.~S.}\ \bibnamefont {Dresselhaus}},\
  }\bibfield  {title} {\enquote {\bibinfo {title} {Third-order intermodulation
  distortion in {YBa$_{2}$}{Cu$_{3}$}{O$_{7-\ensuremath{\delta}}$} grain
  boundaries},}\ }\href {\doibase 10.1103/PhysRevB.65.214533} {\bibfield
  {journal} {\bibinfo  {journal} {Phys. Rev. B}\ }\textbf {\bibinfo {volume}
  {65}},\ \bibinfo {pages} {214533} (\bibinfo {year} {2002})}\BibitemShut
  {NoStop}%
\bibitem [{\citenamefont {Willemsen}\ \emph {et~al.}(1998)\citenamefont
  {Willemsen}, \citenamefont {Kihlstrom}, \citenamefont {Dahm}, \citenamefont
  {Scalapino}, \citenamefont {Gowe}, \citenamefont {Bonn},\ and\ \citenamefont
  {Hardy}}]{Willemsen1998}%
  \BibitemOpen
  \bibfield  {author} {\bibinfo {author} {\bibfnamefont {Balam~A.}\
  \bibnamefont {Willemsen}}, \bibinfo {author} {\bibfnamefont {K.~E.}\
  \bibnamefont {Kihlstrom}}, \bibinfo {author} {\bibfnamefont {T.}~\bibnamefont
  {Dahm}}, \bibinfo {author} {\bibfnamefont {D.~J.}\ \bibnamefont {Scalapino}},
  \bibinfo {author} {\bibfnamefont {B.}~\bibnamefont {Gowe}}, \bibinfo {author}
  {\bibfnamefont {D.~A.}\ \bibnamefont {Bonn}}, \ and\ \bibinfo {author}
  {\bibfnamefont {W.~N.}\ \bibnamefont {Hardy}},\ }\bibfield  {title} {\enquote
  {\bibinfo {title} {Microwave loss and intermodulation in
  {Tl$_{2}$}{Ba$_{2}$}{CaCu$_{2}$}{O$_{y}$} thin films},}\ }\href {\doibase
  10.1103/PhysRevB.58.6650} {\bibfield  {journal} {\bibinfo  {journal} {Phys.
  Rev. B}\ }\textbf {\bibinfo {volume} {58}},\ \bibinfo {pages} {6650--6654}
  (\bibinfo {year} {1998})}\BibitemShut {NoStop}%
\bibitem [{\citenamefont {Hao}\ \emph {et~al.}(2001)\citenamefont {Hao},
  \citenamefont {Gallop}, \citenamefont {Purnell}, \citenamefont {Cohen},\ and\
  \citenamefont {Thiess}}]{Hao2001}%
  \BibitemOpen
  \bibfield  {author} {\bibinfo {author} {\bibfnamefont {Ling}\ \bibnamefont
  {Hao}}, \bibinfo {author} {\bibfnamefont {J.}~\bibnamefont {Gallop}},
  \bibinfo {author} {\bibfnamefont {A.}~\bibnamefont {Purnell}}, \bibinfo
  {author} {\bibfnamefont {L.}~\bibnamefont {Cohen}}, \ and\ \bibinfo {author}
  {\bibfnamefont {S.}~\bibnamefont {Thiess}},\ }\bibfield  {title} {\enquote
  {\bibinfo {title} {Non-linear microwave response of {HTS} thin films: a
  comparison of intermodulation and conventional measurements},}\ }\href
  {\doibase 10.1109/77.919795} {\bibfield  {journal} {\bibinfo  {journal} {IEEE
  Transactions on Applied Superconductivity}\ }\textbf {\bibinfo {volume}
  {11}},\ \bibinfo {pages} {3411--3414} (\bibinfo {year} {2001})}\BibitemShut
  {NoStop}%
\bibitem [{\citenamefont {Lamura}\ \emph {et~al.}(2003)\citenamefont {Lamura},
  \citenamefont {Purnell}, \citenamefont {Cohen}, \citenamefont {Andreone},
  \citenamefont {Chiarella}, \citenamefont {Gennaro}, \citenamefont {Vaglio},
  \citenamefont {Hao},\ and\ \citenamefont {Gallop}}]{Lamura2003}%
  \BibitemOpen
  \bibfield  {author} {\bibinfo {author} {\bibfnamefont {G.}~\bibnamefont
  {Lamura}}, \bibinfo {author} {\bibfnamefont {A.~J.}\ \bibnamefont {Purnell}},
  \bibinfo {author} {\bibfnamefont {L.~F.}\ \bibnamefont {Cohen}}, \bibinfo
  {author} {\bibfnamefont {A.}~\bibnamefont {Andreone}}, \bibinfo {author}
  {\bibfnamefont {F.}~\bibnamefont {Chiarella}}, \bibinfo {author}
  {\bibfnamefont {E.~Di}\ \bibnamefont {Gennaro}}, \bibinfo {author}
  {\bibfnamefont {R.}~\bibnamefont {Vaglio}}, \bibinfo {author} {\bibfnamefont
  {L.}~\bibnamefont {Hao}}, \ and\ \bibinfo {author} {\bibfnamefont
  {J.}~\bibnamefont {Gallop}},\ }\bibfield  {title} {\enquote {\bibinfo {title}
  {Microwave intermodulation distortion of {MgB$_{2}$} thin films},}\
  }\href@noop {} {\bibfield  {journal} {\bibinfo  {journal} {Applied Physics
  Letters}\ }\textbf {\bibinfo {volume} {82}} (\bibinfo {year}
  {2003})}\BibitemShut {NoStop}%
\bibitem [{\citenamefont {Velichko}(2004)}]{Velichko2004}%
  \BibitemOpen
  \bibfield  {author} {\bibinfo {author} {\bibfnamefont {A~V}\ \bibnamefont
  {Velichko}},\ }\bibfield  {title} {\enquote {\bibinfo {title} {Origin of the
  deviation of intermodulation distortion in high{-T$_c$} thin films from the
  classical 3:1 scaling},}\ }\href
  {http://stacks.iop.org/0953-2048/17/i=1/a=001} {\bibfield  {journal}
  {\bibinfo  {journal} {Superconductor Science and Technology}\ }\textbf
  {\bibinfo {volume} {17}},\ \bibinfo {pages} {1} (\bibinfo {year}
  {2004})}\BibitemShut {NoStop}%
\bibitem [{\citenamefont {Oates}\ \emph {et~al.}(2005)\citenamefont {Oates},
  \citenamefont {Park}, \citenamefont {Agassi}, \citenamefont {Koren},\ and\
  \citenamefont {Irgmaier}}]{Oates2005}%
  \BibitemOpen
  \bibfield  {author} {\bibinfo {author} {\bibfnamefont {D.~E.}\ \bibnamefont
  {Oates}}, \bibinfo {author} {\bibfnamefont {S.~H.}\ \bibnamefont {Park}},
  \bibinfo {author} {\bibfnamefont {D.}~\bibnamefont {Agassi}}, \bibinfo
  {author} {\bibfnamefont {G.}~\bibnamefont {Koren}}, \ and\ \bibinfo {author}
  {\bibfnamefont {K.}~\bibnamefont {Irgmaier}},\ }\bibfield  {title} {\enquote
  {\bibinfo {title} {Temperature dependence of intermodulation distortion in
  {YBCO}: understanding nonlinearity},}\ }\href {\doibase
  10.1109/TASC.2005.849367} {\bibfield  {journal} {\bibinfo  {journal} {IEEE
  Transactions on Applied Superconductivity}\ }\textbf {\bibinfo {volume}
  {15}},\ \bibinfo {pages} {3589--3595} (\bibinfo {year} {2005})}\BibitemShut
  {NoStop}%
\bibitem [{\citenamefont {Oates}\ \emph {et~al.}(2007)\citenamefont {Oates},
  \citenamefont {Agassi},\ and\ \citenamefont {Moeckly}}]{Oates2007}%
  \BibitemOpen
  \bibfield  {author} {\bibinfo {author} {\bibfnamefont {D.~E.}\ \bibnamefont
  {Oates}}, \bibinfo {author} {\bibfnamefont {Y.~D.}\ \bibnamefont {Agassi}}, \
  and\ \bibinfo {author} {\bibfnamefont {B.~H.}\ \bibnamefont {Moeckly}},\
  }\bibfield  {title} {\enquote {\bibinfo {title} {Intermodulation distortion
  and nonlinearity in {MgB$_{2}$}: Experiment and theory},}\ }\href {\doibase
  10.1109/TASC.2007.899004} {\bibfield  {journal} {\bibinfo  {journal} {IEEE
  Transactions on Applied Superconductivity}\ }\textbf {\bibinfo {volume}
  {17}},\ \bibinfo {pages} {2871--2874} (\bibinfo {year} {2007})}\BibitemShut
  {NoStop}%
\bibitem [{\citenamefont {Jang}\ \emph {et~al.}(2009)\citenamefont {Jang},
  \citenamefont {Choi}, \citenamefont {Folkman}, \citenamefont {Oates},\ and\
  \citenamefont {Eom}}]{Jang2009}%
  \BibitemOpen
  \bibfield  {author} {\bibinfo {author} {\bibfnamefont {H.~W.}\ \bibnamefont
  {Jang}}, \bibinfo {author} {\bibfnamefont {K.~J.}\ \bibnamefont {Choi}},
  \bibinfo {author} {\bibfnamefont {C.~M.}\ \bibnamefont {Folkman}}, \bibinfo
  {author} {\bibfnamefont {D.~E.}\ \bibnamefont {Oates}}, \ and\ \bibinfo
  {author} {\bibfnamefont {C.~B.}\ \bibnamefont {Eom}},\ }\bibfield  {title}
  {\enquote {\bibinfo {title} {Intermodulation distortion in epitaxial
  {Y-Ba-Cu-O} thick films and multilayers},}\ }\href {\doibase
  10.1109/TASC.2009.2019668} {\bibfield  {journal} {\bibinfo  {journal} {IEEE
  Transactions on Applied Superconductivity}\ }\textbf {\bibinfo {volume}
  {19}},\ \bibinfo {pages} {2855--2858} (\bibinfo {year} {2009})}\BibitemShut
  {NoStop}%
\bibitem [{\citenamefont {Agassi}\ \emph {et~al.}(2009)\citenamefont {Agassi},
  \citenamefont {Oates},\ and\ \citenamefont {Moeckly}}]{Agassi2009}%
  \BibitemOpen
  \bibfield  {author} {\bibinfo {author} {\bibfnamefont {Y.~D.}\ \bibnamefont
  {Agassi}}, \bibinfo {author} {\bibfnamefont {D.~E.}\ \bibnamefont {Oates}}, \
  and\ \bibinfo {author} {\bibfnamefont {B.~H.}\ \bibnamefont {Moeckly}},\
  }\bibfield  {title} {\enquote {\bibinfo {title} {Evidence for non-$s$-wave
  symmetry of the $\ensuremath{\pi}$ gap in {MgB$_{2}$} from intermodulation
  distortion measurements},}\ }\href {\doibase 10.1103/PhysRevB.80.174522}
  {\bibfield  {journal} {\bibinfo  {journal} {Phys. Rev. B}\ }\textbf {\bibinfo
  {volume} {80}},\ \bibinfo {pages} {174522} (\bibinfo {year}
  {2009})}\BibitemShut {NoStop}%
\bibitem [{\citenamefont {Pease}\ \emph {et~al.}(2010)\citenamefont {Pease},
  \citenamefont {Dober},\ and\ \citenamefont {Remillard}}]{Pease2010}%
  \BibitemOpen
  \bibfield  {author} {\bibinfo {author} {\bibfnamefont {Evan~K.}\ \bibnamefont
  {Pease}}, \bibinfo {author} {\bibfnamefont {Bradley~J.}\ \bibnamefont
  {Dober}}, \ and\ \bibinfo {author} {\bibfnamefont {S.~K.}\ \bibnamefont
  {Remillard}},\ }\bibfield  {title} {\enquote {\bibinfo {title} {Synchronous
  measurement of even and odd order intermodulation distortion at the resonant
  frequency of a superconducting resonator},}\ }\href
  {http://scitation.aip.org/content/aip/journal/rsi/81/2/10.1063/1.3301425}
  {\bibfield  {journal} {\bibinfo  {journal} {Review of Scientific
  Instruments}\ }\textbf {\bibinfo {volume} {81}},\ \bibinfo {eid} {024701}
  (\bibinfo {year} {2010})}\BibitemShut {NoStop}%
\bibitem [{\citenamefont {Agassi}\ and\ \citenamefont
  {Oates}(2014)}]{Agassi2014}%
  \BibitemOpen
  \bibfield  {author} {\bibinfo {author} {\bibfnamefont {Y.D.}\ \bibnamefont
  {Agassi}}\ and\ \bibinfo {author} {\bibfnamefont {D.E.}\ \bibnamefont
  {Oates}},\ }\bibfield  {title} {\enquote {\bibinfo {title} {Intermodulation
  distortion and surface resistance in impurity-doped {YBCO} and
  {MgB$_{2}$}},}\ }\href {\doibase
  http://dx.doi.org/10.1016/j.physc.2014.09.006} {\bibfield  {journal}
  {\bibinfo  {journal} {Physica C: Superconductivity and its Applications}\
  }\textbf {\bibinfo {volume} {506}},\ \bibinfo {pages} {119 -- 132} (\bibinfo
  {year} {2014})}\BibitemShut {NoStop}%
\bibitem [{\citenamefont {Abdo}\ \emph {et~al.}(2009)\citenamefont {Abdo},
  \citenamefont {Suchoi}, \citenamefont {Segev}, \citenamefont {Shtempluck},
  \citenamefont {Blencowe},\ and\ \citenamefont {Buks}}]{Abdo2009}%
  \BibitemOpen
  \bibfield  {author} {\bibinfo {author} {\bibfnamefont {B.}~\bibnamefont
  {Abdo}}, \bibinfo {author} {\bibfnamefont {O.}~\bibnamefont {Suchoi}},
  \bibinfo {author} {\bibfnamefont {E.}~\bibnamefont {Segev}}, \bibinfo
  {author} {\bibfnamefont {O.}~\bibnamefont {Shtempluck}}, \bibinfo {author}
  {\bibfnamefont {M.}~\bibnamefont {Blencowe}}, \ and\ \bibinfo {author}
  {\bibfnamefont {E.}~\bibnamefont {Buks}},\ }\bibfield  {title} {\enquote
  {\bibinfo {title} {Intermodulation and parametric amplification in a
  superconducting stripline resonator integrated with a dc-{SQUID}},}\ }\href
  {http://stacks.iop.org/0295-5075/85/i=6/a=68001} {\bibfield  {journal}
  {\bibinfo  {journal} {EPL (Europhysics Letters)}\ }\textbf {\bibinfo {volume}
  {85}},\ \bibinfo {pages} {68001} (\bibinfo {year} {2009})}\BibitemShut
  {NoStop}%
\bibitem [{\citenamefont {Abdo}\ \emph {et~al.}(2006)\citenamefont {Abdo},
  \citenamefont {Segev}, \citenamefont {Shtempluck},\ and\ \citenamefont
  {Buks}}]{Abdo2006}%
  \BibitemOpen
  \bibfield  {author} {\bibinfo {author} {\bibfnamefont {Baleegh}\ \bibnamefont
  {Abdo}}, \bibinfo {author} {\bibfnamefont {Eran}\ \bibnamefont {Segev}},
  \bibinfo {author} {\bibfnamefont {Oleg}\ \bibnamefont {Shtempluck}}, \ and\
  \bibinfo {author} {\bibfnamefont {Eyal}\ \bibnamefont {Buks}},\ }\bibfield
  {title} {\enquote {\bibinfo {title} {Intermodulation gain in nonlinear {NbN}
  superconducting microwave resonators},}\ }\href
  {http://scitation.aip.org/content/aip/journal/apl/88/2/10.1063/1.2164925}
  {\bibfield  {journal} {\bibinfo  {journal} {Applied Physics Letters}\
  }\textbf {\bibinfo {volume} {88}},\ \bibinfo {eid} {022508} (\bibinfo {year}
  {2006})}\BibitemShut {NoStop}%
\bibitem [{\citenamefont {Dahm}\ and\ \citenamefont
  {Scalapino}(1996)}]{Dahm1996}%
  \BibitemOpen
  \bibfield  {author} {\bibinfo {author} {\bibfnamefont {T.}~\bibnamefont
  {Dahm}}\ and\ \bibinfo {author} {\bibfnamefont {D.~J.}\ \bibnamefont
  {Scalapino}},\ }\bibfield  {title} {\enquote {\bibinfo {title} {Theory of
  microwave intermodulation in a high{-T$_c$} superconducting microstrip
  resonator},}\ }\href {\doibase http://dx.doi.org/10.1063/1.116960} {\bibfield
   {journal} {\bibinfo  {journal} {Applied Physics Letters}\ }\textbf {\bibinfo
  {volume} {69}},\ \bibinfo {pages} {4248--4250} (\bibinfo {year}
  {1996})}\BibitemShut {NoStop}%
\bibitem [{\citenamefont {Sollner}\ \emph {et~al.}(1996)\citenamefont
  {Sollner}, \citenamefont {Sage},\ and\ \citenamefont {Oates}}]{Sollner1996}%
  \BibitemOpen
  \bibfield  {author} {\bibinfo {author} {\bibfnamefont {T.~C. L.~Gerhard}\
  \bibnamefont {Sollner}}, \bibinfo {author} {\bibfnamefont {Jay~P.}\
  \bibnamefont {Sage}}, \ and\ \bibinfo {author} {\bibfnamefont {Daniel~E.}\
  \bibnamefont {Oates}},\ }\bibfield  {title} {\enquote {\bibinfo {title}
  {Microwave intermodulation products and excess critical current in
  {YBa$_{2}$Cu$_{3}$O$_{7-x}$} {J}osephson junctions},}\ }\href {\doibase
  http://dx.doi.org/10.1063/1.116209} {\bibfield  {journal} {\bibinfo
  {journal} {Applied Physics Letters}\ }\textbf {\bibinfo {volume} {68}},\
  \bibinfo {pages} {1003--1005} (\bibinfo {year} {1996})}\BibitemShut {NoStop}%
\bibitem [{\citenamefont {Dahm}\ and\ \citenamefont
  {Scalapino}(1997{\natexlab{b}})}]{Dahm1997theoryres}%
  \BibitemOpen
  \bibfield  {author} {\bibinfo {author} {\bibfnamefont {T.}~\bibnamefont
  {Dahm}}\ and\ \bibinfo {author} {\bibfnamefont {D.~J.}\ \bibnamefont
  {Scalapino}},\ }\bibfield  {title} {\enquote {\bibinfo {title} {Theory of
  intermodulation in a superconducting microstrip resonator},}\ }\href
  {\doibase http://dx.doi.org/10.1063/1.364056} {\bibfield  {journal} {\bibinfo
   {journal} {Journal of Applied Physics}\ }\textbf {\bibinfo {volume} {81}},\
  \bibinfo {pages} {2002--2009} (\bibinfo {year}
  {1997}{\natexlab{b}})}\BibitemShut {NoStop}%
\bibitem [{\citenamefont {Willemsen}\ \emph {et~al.}(1999)\citenamefont
  {Willemsen}, \citenamefont {Kihlstrom},\ and\ \citenamefont
  {Dahm}}]{Willemsen1999}%
  \BibitemOpen
  \bibfield  {author} {\bibinfo {author} {\bibfnamefont {Balam~A.}\
  \bibnamefont {Willemsen}}, \bibinfo {author} {\bibfnamefont {K.~E.}\
  \bibnamefont {Kihlstrom}}, \ and\ \bibinfo {author} {\bibfnamefont
  {T.}~\bibnamefont {Dahm}},\ }\bibfield  {title} {\enquote {\bibinfo {title}
  {Unusual power dependence of two-tone intermodulation in high{-T$_c$}
  superconducting microwave resonators},}\ }\href {\doibase
  http://dx.doi.org/10.1063/1.123112} {\bibfield  {journal} {\bibinfo
  {journal} {Applied Physics Letters}\ }\textbf {\bibinfo {volume} {74}},\
  \bibinfo {pages} {753--755} (\bibinfo {year} {1999})}\BibitemShut {NoStop}%
\bibitem [{\citenamefont {Dahm}\ \emph
  {et~al.}(1999{\natexlab{a}})\citenamefont {Dahm}, \citenamefont {Scalapino},\
  and\ \citenamefont {Willemsen}}]{Dahm1999}%
  \BibitemOpen
  \bibfield  {author} {\bibinfo {author} {\bibfnamefont {T.}~\bibnamefont
  {Dahm}}, \bibinfo {author} {\bibfnamefont {D.~J.}\ \bibnamefont {Scalapino}},
  \ and\ \bibinfo {author} {\bibfnamefont {B.~A.}\ \bibnamefont {Willemsen}},\
  }\bibfield  {title} {\enquote {\bibinfo {title} {Phenomenological theory of
  intermodulation in {HTS} resonators and filters},}\ }\href {\doibase
  10.1023/A:1007749215243} {\bibfield  {journal} {\bibinfo  {journal} {Journal
  of Superconductivity}\ }\textbf {\bibinfo {volume} {12}},\ \bibinfo {pages}
  {339--351} (\bibinfo {year} {1999}{\natexlab{a}})}\BibitemShut {NoStop}%
\bibitem [{\citenamefont {Dahm}\ \emph
  {et~al.}(1999{\natexlab{b}})\citenamefont {Dahm}, \citenamefont {Scalapino},\
  and\ \citenamefont {Willemsen}}]{Dahm1999JAP}%
  \BibitemOpen
  \bibfield  {author} {\bibinfo {author} {\bibfnamefont {T.}~\bibnamefont
  {Dahm}}, \bibinfo {author} {\bibfnamefont {D.~J.}\ \bibnamefont {Scalapino}},
  \ and\ \bibinfo {author} {\bibfnamefont {Balam~A.}\ \bibnamefont
  {Willemsen}},\ }\bibfield  {title} {\enquote {\bibinfo {title} {Microwave
  intermodulation of a superconducting disk resonator},}\ }\href {\doibase
  http://dx.doi.org/10.1063/1.371330} {\bibfield  {journal} {\bibinfo
  {journal} {Journal of Applied Physics}\ }\textbf {\bibinfo {volume} {86}},\
  \bibinfo {pages} {4055--4057} (\bibinfo {year}
  {1999}{\natexlab{b}})}\BibitemShut {NoStop}%
\bibitem [{\citenamefont {Hu}\ \emph {et~al.}(1999)\citenamefont {Hu},
  \citenamefont {Thanawalla}, \citenamefont {Feenstra}, \citenamefont
  {Wellstood},\ and\ \citenamefont {Anlage}}]{Hu1999}%
  \BibitemOpen
  \bibfield  {author} {\bibinfo {author} {\bibfnamefont {Wensheng}\
  \bibnamefont {Hu}}, \bibinfo {author} {\bibfnamefont {A.~S.}\ \bibnamefont
  {Thanawalla}}, \bibinfo {author} {\bibfnamefont {B.~J.}\ \bibnamefont
  {Feenstra}}, \bibinfo {author} {\bibfnamefont {F.~C.}\ \bibnamefont
  {Wellstood}}, \ and\ \bibinfo {author} {\bibfnamefont {Steven~M.}\
  \bibnamefont {Anlage}},\ }\bibfield  {title} {\enquote {\bibinfo {title}
  {Imaging of microwave intermodulation fields in a superconducting microstrip
  resonator},}\ }\href {\doibase http://dx.doi.org/10.1063/1.125162} {\bibfield
   {journal} {\bibinfo  {journal} {Applied Physics Letters}\ }\textbf {\bibinfo
  {volume} {75}},\ \bibinfo {pages} {2824--2826} (\bibinfo {year}
  {1999})}\BibitemShut {NoStop}%
\bibitem [{\citenamefont {Vopilkin}\ \emph {et~al.}(2000)\citenamefont
  {Vopilkin}, \citenamefont {Parafin},\ and\ \citenamefont
  {Reznik}}]{Vopilkin2000}%
  \BibitemOpen
  \bibfield  {author} {\bibinfo {author} {\bibfnamefont {E.~A.}\ \bibnamefont
  {Vopilkin}}, \bibinfo {author} {\bibfnamefont {A.~E.}\ \bibnamefont
  {Parafin}}, \ and\ \bibinfo {author} {\bibfnamefont {A.~N.}\ \bibnamefont
  {Reznik}},\ }\bibfield  {title} {\enquote {\bibinfo {title} {Intermodulation
  in a microwave resonator with a high-temperature superconductor},}\ }\href
  {\doibase 10.1134/1.1259600} {\bibfield  {journal} {\bibinfo  {journal}
  {Technical Physics}\ }\textbf {\bibinfo {volume} {45}},\ \bibinfo {pages}
  {214--220} (\bibinfo {year} {2000})}\BibitemShut {NoStop}%
\bibitem [{\citenamefont {Hutter}\ \emph {et~al.}(2010)\citenamefont {Hutter},
  \citenamefont {Platz}, \citenamefont {Thol\'en}, \citenamefont {Hansson},\
  and\ \citenamefont {Haviland}}]{Hutter2010_reconstructIMD}%
  \BibitemOpen
  \bibfield  {author} {\bibinfo {author} {\bibfnamefont {Carsten}\ \bibnamefont
  {Hutter}}, \bibinfo {author} {\bibfnamefont {Daniel}\ \bibnamefont {Platz}},
  \bibinfo {author} {\bibfnamefont {E.~A.}\ \bibnamefont {Thol\'en}}, \bibinfo
  {author} {\bibfnamefont {T.~H.}\ \bibnamefont {Hansson}}, \ and\ \bibinfo
  {author} {\bibfnamefont {D.~B.}\ \bibnamefont {Haviland}},\ }\bibfield
  {title} {\enquote {\bibinfo {title} {Reconstructing nonlinearities with
  intermodulation spectroscopy},}\ }\href {\doibase
  10.1103/PhysRevLett.104.050801} {\bibfield  {journal} {\bibinfo  {journal}
  {Phys. Rev. Lett.}\ }\textbf {\bibinfo {volume} {104}},\ \bibinfo {pages}
  {050801} (\bibinfo {year} {2010})}\BibitemShut {NoStop}%
\bibitem [{\citenamefont {Anlage}(2011)}]{Anlage2011}%
  \BibitemOpen
  \bibfield  {author} {\bibinfo {author} {\bibfnamefont {S.~M.}\ \bibnamefont
  {Anlage}},\ }\bibfield  {title} {\enquote {\bibinfo {title} {The physics and
  applications of superconducting metamaterials},}\ }\href@noop {} {\bibfield
  {journal} {\bibinfo  {journal} {J. Opt.}\ }\textbf {\bibinfo {volume} {13}},\
  \bibinfo {pages} {024001} (\bibinfo {year} {2011})}\BibitemShut {NoStop}%
\bibitem [{\citenamefont {Jung}\ \emph
  {et~al.}(2014{\natexlab{a}})\citenamefont {Jung}, \citenamefont {Ustinov},\
  and\ \citenamefont {Anlage}}]{Jung2014review}%
  \BibitemOpen
  \bibfield  {author} {\bibinfo {author} {\bibfnamefont {Philipp}\ \bibnamefont
  {Jung}}, \bibinfo {author} {\bibfnamefont {Alexey~V}\ \bibnamefont
  {Ustinov}}, \ and\ \bibinfo {author} {\bibfnamefont {Steven~M}\ \bibnamefont
  {Anlage}},\ }\bibfield  {title} {\enquote {\bibinfo {title} {Progress in
  superconducting metamaterials},}\ }\href
  {http://stacks.iop.org/0953-2048/27/i=7/a=073001} {\bibfield  {journal}
  {\bibinfo  {journal} {Superconductor Science and Technology}\ }\textbf
  {\bibinfo {volume} {27}},\ \bibinfo {pages} {073001} (\bibinfo {year}
  {2014}{\natexlab{a}})}\BibitemShut {NoStop}%
\bibitem [{\citenamefont {Jung}\ \emph {et~al.}(2013)\citenamefont {Jung},
  \citenamefont {Butz}, \citenamefont {Shitov},\ and\ \citenamefont
  {Ustinov}}]{Jung2013}%
  \BibitemOpen
  \bibfield  {author} {\bibinfo {author} {\bibfnamefont {P.}~\bibnamefont
  {Jung}}, \bibinfo {author} {\bibfnamefont {S.}~\bibnamefont {Butz}}, \bibinfo
  {author} {\bibfnamefont {S.~V.}\ \bibnamefont {Shitov}}, \ and\ \bibinfo
  {author} {\bibfnamefont {A.~V.}\ \bibnamefont {Ustinov}},\ }\bibfield
  {title} {\enquote {\bibinfo {title} {Low-loss tunable metamaterials using
  superconducting circuits with {J}osephson junctions},}\ }\href@noop {}
  {\bibfield  {journal} {\bibinfo  {journal} {Appl. Phys. Lett.}\ }\textbf
  {\bibinfo {volume} {102}},\ \bibinfo {pages} {062601--4} (\bibinfo {year}
  {2013})}\BibitemShut {NoStop}%
\bibitem [{\citenamefont {Butz}\ \emph {et~al.}(2013)\citenamefont {Butz},
  \citenamefont {Jung}, \citenamefont {Filippenko}, \citenamefont {Koshelets},\
  and\ \citenamefont {Ustinov}}]{Butz20132}%
  \BibitemOpen
  \bibfield  {author} {\bibinfo {author} {\bibfnamefont {S.}~\bibnamefont
  {Butz}}, \bibinfo {author} {\bibfnamefont {P.}~\bibnamefont {Jung}}, \bibinfo
  {author} {\bibfnamefont {L.~V.}\ \bibnamefont {Filippenko}}, \bibinfo
  {author} {\bibfnamefont {V.~P.}\ \bibnamefont {Koshelets}}, \ and\ \bibinfo
  {author} {\bibfnamefont {A.~V.}\ \bibnamefont {Ustinov}},\ }\bibfield
  {title} {\enquote {\bibinfo {title} {A one-dimensional tunable magnetic
  metamaterial},}\ }\href@noop {} {\bibfield  {journal} {\bibinfo  {journal}
  {Opt. Express}\ }\textbf {\bibinfo {volume} {21}},\ \bibinfo {pages}
  {22540--22548} (\bibinfo {year} {2013})}\BibitemShut {NoStop}%
\bibitem [{\citenamefont {Trepanier}\ \emph {et~al.}(2013)\citenamefont
  {Trepanier}, \citenamefont {Zhang}, \citenamefont {Mukhanov},\ and\
  \citenamefont {Anlage}}]{Trepanier2013}%
  \BibitemOpen
  \bibfield  {author} {\bibinfo {author} {\bibfnamefont {M.}~\bibnamefont
  {Trepanier}}, \bibinfo {author} {\bibfnamefont {Daimeng}\ \bibnamefont
  {Zhang}}, \bibinfo {author} {\bibfnamefont {Oleg}\ \bibnamefont {Mukhanov}},
  \ and\ \bibinfo {author} {\bibfnamefont {Steven~M.}\ \bibnamefont {Anlage}},\
  }\bibfield  {title} {\enquote {\bibinfo {title} {Realization and modeling of
  metamaterials made of rf superconducting quantum-interference devices},}\
  }\href {\doibase 10.1103/PhysRevX.3.041029} {\bibfield  {journal} {\bibinfo
  {journal} {Phys. Rev. X}\ }\textbf {\bibinfo {volume} {3}},\ \bibinfo {pages}
  {041029} (\bibinfo {year} {2013})}\BibitemShut {NoStop}%
\bibitem [{\citenamefont {Zhang}\ \emph {et~al.}(2015)\citenamefont {Zhang},
  \citenamefont {Trepanier}, \citenamefont {Mukhanov},\ and\ \citenamefont
  {Anlage}}]{Daimeng2015}%
  \BibitemOpen
  \bibfield  {author} {\bibinfo {author} {\bibfnamefont {Daimeng}\ \bibnamefont
  {Zhang}}, \bibinfo {author} {\bibfnamefont {Melissa}\ \bibnamefont
  {Trepanier}}, \bibinfo {author} {\bibfnamefont {Oleg}\ \bibnamefont
  {Mukhanov}}, \ and\ \bibinfo {author} {\bibfnamefont {Steven~M.}\
  \bibnamefont {Anlage}},\ }\bibfield  {title} {\enquote {\bibinfo {title}
  {Tunable broadband transparency of macroscopic quantum superconducting
  metamaterials},}\ }\href {\doibase 10.1103/PhysRevX.5.041045} {\bibfield
  {journal} {\bibinfo  {journal} {Phys. Rev. X}\ }\textbf {\bibinfo {volume}
  {5}},\ \bibinfo {pages} {041045} (\bibinfo {year} {2015})}\BibitemShut
  {NoStop}%
\bibitem [{\citenamefont {Mukhanov}\ \emph {et~al.}(2014)\citenamefont
  {Mukhanov}, \citenamefont {Prokopenko},\ and\ \citenamefont
  {Romanofsky}}]{Mukhanov2014}%
  \BibitemOpen
  \bibfield  {author} {\bibinfo {author} {\bibfnamefont {O.}~\bibnamefont
  {Mukhanov}}, \bibinfo {author} {\bibfnamefont {G.}~\bibnamefont
  {Prokopenko}}, \ and\ \bibinfo {author} {\bibfnamefont {R.}~\bibnamefont
  {Romanofsky}},\ }\bibfield  {title} {\enquote {\bibinfo {title} {Quantum
  sensitivity: Superconducting quantum interference filter-based microwave
  receivers},}\ }\href {\doibase 10.1109/MMM.2014.2332421} {\bibfield
  {journal} {\bibinfo  {journal} {IEEE Microwave Magazine}\ }\textbf {\bibinfo
  {volume} {15}},\ \bibinfo {pages} {57--65} (\bibinfo {year}
  {2014})}\BibitemShut {NoStop}%
\bibitem [{\citenamefont {Mukhanov}\ \emph {et~al.}(2008)\citenamefont
  {Mukhanov}, \citenamefont {Kirichenko}, \citenamefont {Vernik}, \citenamefont
  {Filippov}, \citenamefont {Kirichenko}, \citenamefont {Webber}, \citenamefont
  {Dotsenko}, \citenamefont {Talalaevskii}, \citenamefont {Tang}, \citenamefont
  {Sahu}, \citenamefont {Shevchenko}, \citenamefont {Miller}, \citenamefont
  {Kaplan}, \citenamefont {Sarwana},\ and\ \citenamefont
  {Gupta}}]{Mukhanov2008}%
  \BibitemOpen
  \bibfield  {author} {\bibinfo {author} {\bibfnamefont {O.~A.}\ \bibnamefont
  {Mukhanov}}, \bibinfo {author} {\bibfnamefont {D.}~\bibnamefont
  {Kirichenko}}, \bibinfo {author} {\bibfnamefont {I.~V.}\ \bibnamefont
  {Vernik}}, \bibinfo {author} {\bibfnamefont {T.~V.}\ \bibnamefont
  {Filippov}}, \bibinfo {author} {\bibfnamefont {A.}~\bibnamefont
  {Kirichenko}}, \bibinfo {author} {\bibfnamefont {R.}~\bibnamefont {Webber}},
  \bibinfo {author} {\bibfnamefont {V.}~\bibnamefont {Dotsenko}}, \bibinfo
  {author} {\bibfnamefont {A.}~\bibnamefont {Talalaevskii}}, \bibinfo {author}
  {\bibfnamefont {J.~C.}\ \bibnamefont {Tang}}, \bibinfo {author}
  {\bibfnamefont {A.}~\bibnamefont {Sahu}}, \bibinfo {author} {\bibfnamefont
  {P.}~\bibnamefont {Shevchenko}}, \bibinfo {author} {\bibfnamefont
  {R.}~\bibnamefont {Miller}}, \bibinfo {author} {\bibfnamefont {S.~B.}\
  \bibnamefont {Kaplan}}, \bibinfo {author} {\bibfnamefont {S.}~\bibnamefont
  {Sarwana}}, \ and\ \bibinfo {author} {\bibfnamefont {D.}~\bibnamefont
  {Gupta}},\ }\bibfield  {title} {\enquote {\bibinfo {title} {Superconductor
  digital-{RF} receiver systems},}\ }\href@noop {} {\bibfield  {journal}
  {\bibinfo  {journal} {IEICE Trans. Electron.}\ }\textbf {\bibinfo {volume}
  {E91-C}},\ \bibinfo {pages} {306--317} (\bibinfo {year} {2008})}\BibitemShut
  {NoStop}%
\bibitem [{\citenamefont {Du}\ \emph {et~al.}(2006)\citenamefont {Du},
  \citenamefont {Chen},\ and\ \citenamefont {Li}}]{Du2006}%
  \BibitemOpen
  \bibfield  {author} {\bibinfo {author} {\bibfnamefont {C.~G.}\ \bibnamefont
  {Du}}, \bibinfo {author} {\bibfnamefont {H.~Y.}\ \bibnamefont {Chen}}, \ and\
  \bibinfo {author} {\bibfnamefont {S.~Q.}\ \bibnamefont {Li}},\ }\bibfield
  {title} {\enquote {\bibinfo {title} {Quantum left-handed metamaterial from
  superconducting quantum-interference devices},}\ }\href@noop {} {\bibfield
  {journal} {\bibinfo  {journal} {Phys. Rev. B}\ }\textbf {\bibinfo {volume}
  {74}},\ \bibinfo {pages} {113105} (\bibinfo {year} {2006})}\BibitemShut
  {NoStop}%
\bibitem [{\citenamefont {Lazarides}\ and\ \citenamefont
  {Tsironis}(2007)}]{Lazarides2007}%
  \BibitemOpen
  \bibfield  {author} {\bibinfo {author} {\bibfnamefont {N.}~\bibnamefont
  {Lazarides}}\ and\ \bibinfo {author} {\bibfnamefont {G.~P.}\ \bibnamefont
  {Tsironis}},\ }\bibfield  {title} {\enquote {\bibinfo {title} {{RF}
  superconducting quantum interference device metamaterials},}\ }\href@noop {}
  {\bibfield  {journal} {\bibinfo  {journal} {Appl. Phys. Lett.}\ }\textbf
  {\bibinfo {volume} {90}},\ \bibinfo {pages} {163501} (\bibinfo {year}
  {2007})}\BibitemShut {NoStop}%
\bibitem [{\citenamefont {Maimistov}\ and\ \citenamefont
  {Gabitov}(2010)}]{Maimistov2010}%
  \BibitemOpen
  \bibfield  {author} {\bibinfo {author} {\bibfnamefont {A.~I.}\ \bibnamefont
  {Maimistov}}\ and\ \bibinfo {author} {\bibfnamefont {I.~R.}\ \bibnamefont
  {Gabitov}},\ }\bibfield  {title} {\enquote {\bibinfo {title} {Nonlinear
  response of a thin metamaterial film containing {J}osephson junctions},}\
  }\href@noop {} {\bibfield  {journal} {\bibinfo  {journal} {Opt. Commun.}\
  }\textbf {\bibinfo {volume} {283}},\ \bibinfo {pages} {1633--1639} (\bibinfo
  {year} {2010})}\BibitemShut {NoStop}%
\bibitem [{\citenamefont {Lazarides}\ and\ \citenamefont
  {Tsironis}(2013)}]{Lazarides2013}%
  \BibitemOpen
  \bibfield  {author} {\bibinfo {author} {\bibfnamefont {N.}~\bibnamefont
  {Lazarides}}\ and\ \bibinfo {author} {\bibfnamefont {G.~P.}\ \bibnamefont
  {Tsironis}},\ }\bibfield  {title} {\enquote {\bibinfo {title} {Multistability
  and self-organization in disordered {SQUID} metamaterials},}\ }\href@noop {}
  {\bibfield  {journal} {\bibinfo  {journal} {Supercond. Sci. Technol.}\
  }\textbf {\bibinfo {volume} {26}},\ \bibinfo {pages} {084006} (\bibinfo
  {year} {2013})}\BibitemShut {NoStop}%
\bibitem [{\citenamefont {Jung}\ \emph
  {et~al.}(2014{\natexlab{b}})\citenamefont {Jung}, \citenamefont {Butz},
  \citenamefont {Koshelets},\ and\ \citenamefont {Ustinov}}]{Jung2014}%
  \BibitemOpen
  \bibfield  {author} {\bibinfo {author} {\bibfnamefont {P.}~\bibnamefont
  {Jung}}, \bibinfo {author} {\bibfnamefont {M.and Fistul M. V.and
  Lepp\"{a}kangas~J.}\ \bibnamefont {Butz}, \bibfnamefont {S.and~Marthaler}},
  \bibinfo {author} {\bibfnamefont {V.~P.}\ \bibnamefont {Koshelets}}, \ and\
  \bibinfo {author} {\bibfnamefont {A.~V.}\ \bibnamefont {Ustinov}},\
  }\bibfield  {title} {\enquote {\bibinfo {title} {Multistability and switching
  in a superconducting metamaterial},}\ }\href@noop {} {\bibfield  {journal}
  {\bibinfo  {journal} {Nat. Comms.}\ }\textbf {\bibinfo {volume} {5}},\
  \bibinfo {pages} {4730} (\bibinfo {year} {2014}{\natexlab{b}})}\BibitemShut
  {NoStop}%
\bibitem [{\citenamefont {Tsironis}\ \emph {et~al.}(2014)\citenamefont
  {Tsironis}, \citenamefont {Lazarides},\ and\ \citenamefont
  {Margaris}}]{Tsironis2014}%
  \BibitemOpen
  \bibfield  {author} {\bibinfo {author} {\bibfnamefont {G.P.}\ \bibnamefont
  {Tsironis}}, \bibinfo {author} {\bibfnamefont {N.}~\bibnamefont {Lazarides}},
  \ and\ \bibinfo {author} {\bibfnamefont {I.}~\bibnamefont {Margaris}},\
  }\bibfield  {title} {\enquote {\bibinfo {title} {Wide-band tuneability,
  nonlinear transmission, and dynamic multistability in {SQUID}
  metamaterials},}\ }\href {\doibase 10.1007/s00339-014-8706-7} {\bibfield
  {journal} {\bibinfo  {journal} {Applied Physics A}\ }\textbf {\bibinfo
  {volume} {117}},\ \bibinfo {pages} {579--588} (\bibinfo {year}
  {2014})}\BibitemShut {NoStop}%
\bibitem [{\citenamefont {Jung}(2014)}]{Jungthesis}%
  \BibitemOpen
  \bibfield  {author} {\bibinfo {author} {\bibfnamefont {Philipp}\ \bibnamefont
  {Jung}},\ }\emph {\bibinfo {title} {Nonlinear Effects in Superconducting
  Quantum Interference Meta-atoms}},\ \href@noop {} {Ph.D. thesis},\ \bibinfo
  {school} {Karlsruher Institut für Technologie (KIT)}, \bibinfo {address}
  {http://www.ksp.kit.edu/download/1000043835} (\bibinfo {year}
  {2014})\BibitemShut {NoStop}%
\bibitem [{\citenamefont {Hizanidis}\ \emph {et~al.}(2016)\citenamefont
  {Hizanidis}, \citenamefont {Lazarides},\ and\ \citenamefont
  {Tsironis}}]{Tsironis2016}%
  \BibitemOpen
  \bibfield  {author} {\bibinfo {author} {\bibfnamefont {J.}~\bibnamefont
  {Hizanidis}}, \bibinfo {author} {\bibfnamefont {N.}~\bibnamefont
  {Lazarides}}, \ and\ \bibinfo {author} {\bibfnamefont {G.~P.}\ \bibnamefont
  {Tsironis}},\ }\href@noop {} {\enquote {\bibinfo {title} {Chimeras in locally
  coupled {SQUIDs}: Lions, goats and snakes},}\ } (\bibinfo {year} {2016}),\
  \Eprint {http://arxiv.org/abs/{arXiv:1604.08160}} {{arXiv:1604.08160}}
  \BibitemShut {NoStop}%
\bibitem [{\citenamefont {Chesca}(1998)}]{Chesca1998}%
  \BibitemOpen
  \bibfield  {author} {\bibinfo {author} {\bibfnamefont {B.}~\bibnamefont
  {Chesca}},\ }\bibfield  {title} {\enquote {\bibinfo {title} {Theory of {RF
  SQUIDs} operating in the presence of large thermal fluctations},}\
  }\href@noop {} {\bibfield  {journal} {\bibinfo  {journal} {J. Low Temp.
  Phys.}\ }\textbf {\bibinfo {volume} {110}},\ \bibinfo {pages} {963--1001}
  (\bibinfo {year} {1998})}\BibitemShut {NoStop}%
\bibitem [{\citenamefont {Likharev}(1986)}]{Likharev1986}%
  \BibitemOpen
  \bibfield  {author} {\bibinfo {author} {\bibfnamefont {K.~K.}\ \bibnamefont
  {Likharev}},\ }\href@noop {} {\emph {\bibinfo {title} {Dynamics of
  {J}osephson Junctions and Circuits}}}\ (\bibinfo  {publisher} {Gordon and
  Breach},\ \bibinfo {address} {New York},\ \bibinfo {year} {1986})\BibitemShut
  {NoStop}%
\bibitem [{\citenamefont {Research}()}]{Fasthenry}%
  \BibitemOpen
  \bibfield  {author} {\bibinfo {author} {\bibfnamefont {Whiteley}\
  \bibnamefont {Research}},\ }\href {http://www.wrcad.com/} {}\bibinfo
  {howpublished} {\url{http://www.wrcad.com/}}\BibitemShut {NoStop}%
\bibitem [{\citenamefont {Doyle}(2008)}]{Doyle2008}%
  \BibitemOpen
  \bibfield  {author} {\bibinfo {author} {\bibfnamefont {S.}~\bibnamefont
  {Doyle}},\ }\bibfield  {title} {\enquote {\bibinfo {title} {Lumped element
  kinetic inductance detectors},}\ }\href@noop {} {\bibfield  {journal}
  {\bibinfo  {journal} {Ph.D. thesis, University of Cardiff, Cardiff, UK}\ }
  (\bibinfo {year} {2008})}\BibitemShut {NoStop}%
\end{thebibliography}%
\end{document}